\documentclass[journal,10pt]{IEEEtran}
\usepackage{amsfonts,amssymb}
\usepackage{amsmath}
\usepackage{array}
\usepackage{cite}
\usepackage{url}
\usepackage{xcolor}
\usepackage{cite,graphicx,amsmath,amssymb}
\usepackage{subfigure}
\usepackage{fancyhdr}
\usepackage{mdwmath}
\usepackage{mdwtab}
\usepackage{amsthm}
\usepackage{booktabs}
\usepackage{multirow}
\usepackage{longtable}
\usepackage{rotating}
\usepackage{makecell}

\newtheorem{remark}{Remark}
\newtheorem{theorem}{Theorem}
\newtheorem{definition}{Definition}

\newtheorem{lemma}{Lemma}

\newtheorem{corollary}{Corollary}

\newtheorem{proposition}{Proposition}

\usepackage[breaklinks,colorlinks,linkcolor=black,citecolor=black,urlcolor=black]{hyperref}

\begin{document}

\title{Physical Layer Security for STAR-RIS-NOMA: A Stochastic Geometry Approach}

\author{Ziyi~Xie,~\IEEEmembership{Graduate Student Member,~IEEE,}
Yuanwei~Liu,~\IEEEmembership{Senior Member,~IEEE,} Wenqiang~Yi,~\IEEEmembership{Member,~IEEE,}
Xuanli~Wu,~\IEEEmembership{Member,~IEEE,}  and Arumugam~Nallanathan,~\IEEEmembership{Fellow,~IEEE}

	
\thanks{Z. Xie and X. Wu are with the School of Electronics and Information Engineering, Harbin Institute of Technology, Harbin 150001, China (email: \{ziyi.xie, xlwu2002\}@hit.edu.cn).}

\thanks{Y. Liu and A. Nallanathan are with the School of Electronic Engineering and Computer Science, Queen Mary University of London, E1 4NS, U.K. (email: \{yuanwei.liu, a.nallanathan\}@qmul.ac.uk).}

\thanks{W. Yi is with the School of Computer Science and Electronic Engineering, University of Essex, Colchester CO4 3SQ, U.K. (email: wy23627@essex.ac.uk).}

\thanks{Part of this work has been accepted in IEEE Global Communications Conference (GLOBECOM), 2023 \cite{0conference}.}
  }



\maketitle

\begin{abstract}
In this paper, a stochastic geometry based analytical framework is proposed for secure simultaneous transmitting and reflecting reconfigurable intelligent surface (STAR-RIS) assisted non-orthogonal multiple access (NOMA) transmissions, where legitimate users (LUs) and eavesdroppers are randomly distributed. Both the time-switching protocol (TS) and energy splitting (ES) protocol are considered for the STAR-RIS. To characterize system performance, the channel statistics are first provided, and the Gamma approximation is adopted for general cascaded $\kappa$-$\mu$ fading. Afterward, the closed-form expressions for both the secrecy outage probability (SOP) and average secrecy capacity (ASC) are derived. To obtain further insights, the asymptotic performance for the secrecy diversity order and the secrecy slope are deduced. The theoretical results show that 1) the secrecy diversity orders of the strong LU and the weak LU depend on the path loss exponent and the distribution of the received signal-to-noise ratio, respectively; 2) the secrecy slope of the ES protocol achieves the value of one, higher than the slope of the TS protocol which is the mode operation parameter of TS. The numerical results demonstrate that: 1) there is an optimal STAR-RIS mode operation parameter to maximize the secrecy performance; 2) the STAR-RIS-NOMA significantly outperforms the STAR-RIS-orthogonal multiple access.
\end{abstract}


\section{Introduction}
Reconfigurable intelligent surfaces (RISs) have been regarded as a promising technique to support the smart radio environment and efficient secure transmissions in future communication networks\cite{RISSecu0,0surveyMarco,0surveyLiu}. One typical RIS is a uniform planer array with a large number of low-cost elements.  By equipping with advanced beamforming controllers, the phase shifts of reflected signals on each RIS element can be changed independently, which helps to adjust the propagation of signals \cite{RISimp0}. Benefiting from this feature, the RIS is able to improve the communication quality of legitimate users (LUs) while limiting eavesdropping by appropriate design on beamforming \cite{RISsurel0}, thereby enhancing physical layer security (PLS). 

For the conventional reflecting-only RIS, PLS performance within half of the space in front of the RIS can be controlled, while the LUs in another half of the space still suffer from uncontrollable eavesdropping. To address this issue, the concept of simultaneous transmitting and reflecting RIS (STAR-RIS) has been proposed \cite{STARmag0,STARJ0}. With three operation protocols, i.e., time switching (TS), energy splitting (ES), and mode switching \cite{STARmag0}, different beamforming approaches can be implemented at both sides of the STAR-RIS, and hence the full-space PLS enhancement is realized. Note that STAR-RISs serve LUs at different sides by the same signal source, a multiple access scheme is indispensable for splitting unicast reflected and transmitted signals. The non-orthogonal multiple access (NOMA) scheme can be a competent candidate due to its high spectral efficiency and user fairness. By employing the superposition coding at the transmitter for power multiplexing and the successive interference cancellation (SIC) at the receiver for detection, STAR-RIS-NOMA protects multiple LUs within the same time-frequency resource block\cite{NOMA10,NOMA20}. 


  
\subsection{Related Works}
Due to the broadcast nature of wireless communications, the concept of PLS was proposed from an information-theoretical perspective \cite{PLSconcept} and has attracted wide attention in recent years. In general, PLS leverages the inherent characteristics of the propagation environment (e.g., fading, noise, and interference) to provide secure transmissions. One common method is to employ the jamming and artificial noise (AN) aided technique to depress the wiretapping of the potential eavesdroppers (Eves) \cite{PLSAN1,PLSAN2}. By deploying jammers that emit jamming signals to confuse the Eves, the information loss due to eavesdropping can be reduced.
Another popular method is to improve the received signal quality at the LUs and to reduce the information leakage to the Eves with the aid of multi-antenna technology \cite{PLSbeam1,PLSbeam2}. Therefore, it is natural to use the RIS which is a passive multi-antenna device for PLS enhancement. The authors of \cite{RISPLS1} focused on a challenging case in downlink RIS-assisted secure transmission, where the Eve has better channel conditions than the LU. The design of beamforming is based on the global channel state information (CSI) of the Eve and the LU. In \cite{RISPLS0}, the authors proposed a novel design on RIS beamforming to eliminate the signals received by the Eve, and hence the global CSI of the Eve is required.
In \cite{RISPLS2}, the secrecy data rate in a RIS-aided massive multiple-input multiple-output system was studied and the statistical CSI of the Eve was considered. These works assumed that the perfect CSI of the Eve is known for beamforming design. In practice, however, it is difficult to acquire perfect knowledge of the CSI of the eavesdropping channels because potential Eves are not continuously communicating with the BS and are even passive to hide their existence. In \cite{RISPLSiSIC1}, the authors utilized the imperfect CSI of the Eves to jointly design the transmit beamformers, AN covariance matrix, and RIS phase shifters. In \cite{RISPLSiSIC2}, the theoretical SOP was derived under the assumption that Eve's CSI is unknown. The above works investigated PLS in the presence of fixed LUs and Eves. To capture the randomness property in the considered space, stochastic geometry is a powerful tool \cite{GPP} and has been widely utilized to study the PLS in traditional communication systems \cite{LiuSOP,SCPLS}. The authors in \cite{RISPLSiSIC3} considered the spatial effect in a RIS-assisted multiple-input multiple-output system and modeled the locations of LUs by a homogeneous Poisson point process (HPPP). Then the theoretical secrecy performance expressions were derived.

Sparked by the potential advantages of combining STAR-RIS and NOMA, research contributions have been devoted to STAR-RIS-NOMA recently.  
In \cite{STARSNOMA1}, the authors focused on the coverage performance and illustrated the superiority of NOMA over orthogonal multiple access (OMA) in STAR-RIS-aided transmissions. In \cite{STARSNOMA2}, bit error rate expressions were derived in STAR-RIS-NOMA and the results revealed that the STAR-RIS-NOMA outperforms the classical NOMA system in terms of error performance. The authors in \cite{STARNOMA0} solved a joint optimization problem for maximizing the achievable sum rate. Simulation results demonstrated the better performance of STAR-RIS-NOMA than the conventional RIS-aided transmissions. Works \cite{STARSNOMA3} and \cite{STARchannel} investigated the theoretical performance of STAR-RIS-NOMA in large-scale networks. In \cite{STARSNOMA3}, the authors proved that for three STAR-RIS operation protocols, the accurate diversity orders depend on the number of STAR-RIS elements that form the passive beamforming. In \cite{STARchannel}, a general analytical framework was provided for the multi-cell networks, where LUs, BSs, and STAR-RISs are randomly distributed. Furthermore, a few works started to pay attention to the PLS in STAR-RIS-NOMA. In \cite{STARPLS1}, the authors focused on the AN-assisted downlink transmission for the improved secrecy rate. In \cite{STARPLS}, residual hardware impairments were considered and analytical expressions of the SOP were provided for the paired NOMA LUs. 
In \cite{RISPLS3}, the authors aimed to maximize the minimum secrecy capacity in STAR-RIS-aided uplink NOMA networks by joint secrecy beamforming design. However, these initial works considered simplified settings with fixed eavesdropping as the location of the Eve is predefined. 

\vspace{-0.3 cm}
\subsection{Motivations and Contributions}
As we have discussed, although STAR-RIS-NOMA has the capability of providing full-space security enhancement, the impact of eavesdropping from the full space needs to be investigated. The security performance of STAR-RIS-NOMA transmissions considering randomly distributed Eves in the full space is important but has not been studied in previous work to our knowledge. Motivated by the above, in this work, we focus on the theoretical security performance of the STAR-RIS-NOMA in the presence of randomly distributed LUs and  Eves.\footnote{In this work, we only consider the impact of external Eves.} The main contributions are summarized as follows:
\begin{itemize}
	\item We propose an analytical framework for STAR-RIS-NOMA in terms of PLS, where both LUs and Eves are randomly deployed. In this framework, the paired NOMA LUs are randomly selected from two sides of the STAR-RIS, and the distribution of Eves is modeled by a HPPP. A mapping method is introduced for unifying the performance at different sides of the STAR-RIS. The beamforming of the STAR-RIS is designed to enhance the channels of LUs. Moreover, a general $\kappa$-$\mu$ distribution is used to characterize the small-scale fading.
	\item We derive the analytical expressions of the SOP for the NOMA LUs when the CSI of Eves is unavailable at the BS. The ordering channel statistics are obtained by exploiting the Gamma distribution to fit the cascaded small-scale fading of STAR-RIS-aided links. We further derive the asymptotic SOP expressions in the high signal-to-noise-ratio (SNR) regime. The analytical results show that the secrecy diversity order for the strong LU is related to the path loss exponent while the error floor exists for the weak LU in the considered scenario.
	\item We derive the analytical expressions of the average secrecy capacity (ASC) performance for the NOMA LUs when the CSI of Eves is available at the BS. The asymptotic ASC is also derived to obtain the secrecy slope. The analytical results demonstrate that the secrecy slopes for the ES protocol and the TS protocol are one and the mode operation parameter of TS, respectively, and hence the ES protocol outperforms the TS protocol in ASC performance when the SNR is high.
	\item We use the numerical results to validate the analysis and to show that: 1) there is an optimal STAR-RIS mode operation parameter to maximize the SOP performance and the ASC; 2) the secrecy performance of the ES protocol always outperforms that of the TS protocol in the considered system; 3) NOMA is able to achieve the higher ASC than the OMA in the STAR-RIS-assisted transmission.
\end{itemize}

\vspace{-0.4 cm}
\subsection{Organization and Notations}
The remainder of this paper is organized as follows. In Section II,
the system model of the secure STAR-RIS-aided NOMA networks is introduced. In Section III, we derive the theoretical SOP for the pair of NOMA LUs, and then the secrecy diversity order is investigated. 
In section IV, we derive the theoretical ASC and then obtain the secrecy slope. 
The numerical results are presented in Section V. Finally, we draw the conclusions in Section VI.

{\it Notation:} $(\cdot)^T$ denotes the transpose operation. $|x|$ is the amplitude of $x$. $\mathbb{E}[\cdot]$ denotes the expectation operator. ${\rm Gamma}(k,\theta)$ is the Gamma distribution with shape $k$ and scale $\theta$.  $\Gamma(x) = \int_0^{\infty} t^{x-1} e^{-t} dt$ is the Gamma function. $\gamma(\alpha,x)$ is the lower incomplete Gamma function \cite[eq. (8.350.1)]{Intetable}. $K_t(x)$ represents is the $t$th-order modified Bessel function of the second kind \cite[eq. (8.432)]{Intetable}. ${}_pF_q({\bf a}_p ; {\bf b}_q; x)$ denotes the generalized hypergeometric function \cite[eq. (9.14.1)]{Intetable}. We denote $[x]^+ = \max \{x,0\}$. $G_{p,q}^{m,n}\left((\cdot) \left| \begin{matrix}
	({\bf a}_p)\\ ({\bf b}_q)
\end{matrix} \right. \right)$ is the Meijer G-function \cite[eq. (9.301)]{Intetable}. For a cumulative distribution function (CDF) $F(x)$, we denote its complementary CDF as ${\bar F}(x) = 1- F(x)$.

\section{System Model}
\begin{figure*}[t!]
	\centering
	\includegraphics[width= 3.6 in]{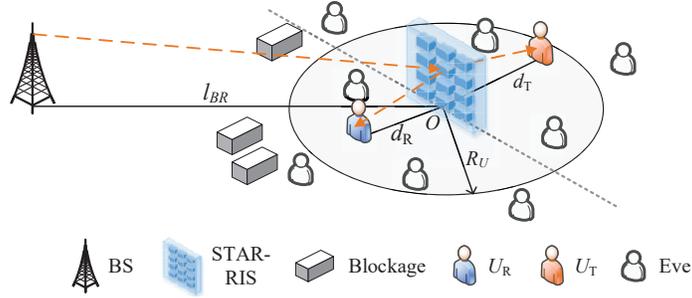}
	\caption{System model for downlink secure STAR-RIS-NOMA transmissions.  }\label{fig1}
\end{figure*}

As shown in Fig. \ref{fig1}, we consider a secure downlink transmission scenario, where a BS communicates with LUs assisted by a STAR-RIS in the presence of Eves. The STAR-RIS with a random orientation is fixed at the origin of a two-dimensional plane $\mathbb{R}^2$. We fix the BS at $(-l_{BR},0)$, while the locations of LUs obey a HPPP ${\bf \Phi}_u$ within a disc area with radius $R_U$ centered at the origin. The spatial distribution of Eves is modeled by another HPPP ${\bf \Phi}_e$ with the density $\lambda_e$ in the considered plane.\footnote{The HPPP is a uniform distribution in this space, which can be regarded as the steady-state distribution in the random direction mobility model \cite{mobility}. Therefore, the LUs and Eves are moving as the random direction model.} We consider that the BS, LUs, and Eves are equipped with a single antenna. The STAR-RIS consists of $N$ elements, and these elements are capable of simultaneously transmitting and reflecting signals.

Those LUs located at the same side of the STAR-RIS as the BS are the reflecting LUs; otherwise the transmitting LUs. We randomly select a reflecting LU $U_{\rm R}$ and a transmitting LU $U_{\rm T}$ to form a typical LU pair. The NOMA transmission scheme is invoked for the typical LU pair.
All Eves have powerful detection capabilities and are able to overhear the messages of all available resource blocks. Moreover, multiuser detection techniques are adopted at Eves, and the Eves can distinguish signals of different LUs when applying the NOMA scheme.
For tractable theoretical expressions, simple setups are employed, and all assumptions are concluded as follows.

{\it Assumption 1:} We consider an urban environment for the secure STAR-RIS-NOMA transmission, so all direct transmission links between the BS and LUs/Eves are blocked.

{\it Assumption 2:} The STAR-RIS is antenna empowered and has the capability of independently controlling the transmitted and reflected signals \cite{ImplementSTAR}.

{\it Assumption 3:} The STAR-RIS is a planar antenna array. We assume that the inter-antennas spacing is volume-unlimited, and the angle difference of the signal transmitted and received can be neglected. In this case, we approximately regard that channel gains of $N$ different channels are independently and identically distributed \cite{Independentc}.

\subsection{Channel Model}
In this work, we mainly focus on the STAR-RIS-aided links between the BS and LUs/Eves. The channel model of the considered STAR-RIS-aided transmission includes the path loss model and the small-scale fading model. For clarity, we use the subscript $\varepsilon = \{{\rm T},{\rm R}\}$ to denote the transmitting LU and the reflecting LU, respectively. 
For LUs, the path loss of the STAR-RIS-aided link is related to the product of two distances, which can be expressed as
\begin{align}
	L_\varepsilon = C_r \left(l_{BR} d_\varepsilon \right)^{-\alpha} ,
\end{align} 
where $d_\varepsilon$ is the distance between the STAR-RIS and the LU. $C_r$ is the reference distance based intercept. $\alpha$ refers to the path loss exponent. Similarly, the path loss of the Eve $i \in {\bf\Phi}_e$ is
\begin{align}
	L_{e, i} = C_r \left(l_{BR} d_e \right)^{-\alpha} .
\end{align}

As in previous works, all channels of the STAR-RIS-aided transmission suffer from cascaded small-scale fading. Specifically, we denote the small-scale fading vectors of the BS-RIS link and the RIS-LU/Eve link as ${\bf h}_{r_1} = [h_{r_1,1},...,h_{r_1,N}]^T$ and ${\bf h}_{r_2} = [h_{r_2, 1},...,h_{r_2, N}]^T$, respectively. For LUs, the power of the equivalent overall small-scale fading for the STAR-RIS-aided cascaded channel is given by $|{h}_{\varepsilon}|^2 = \left| {{\bf h}_{r_2}}^T{\tilde{\bf \Theta}}_\varepsilon  {\bf h}_{r_1} \right|^2$, where ${\tilde{\bf \Theta}}_\varepsilon = {\rm diag} \left(e^{j\theta_{\varepsilon,1}},...,e^{j\theta_{\varepsilon,N}} \right)$ is the normalized phase-shifting matrix of the STAR-RIS, where $j = \sqrt{-1}$ and $\theta_{\varepsilon,n} \in [0,2\pi)$ for $n \in \{1,...,N\}$.
To maximize the received signal power at LUs, the STAR-RIS reconfigures the phase shifts according to the instantaneous exact CSI so that phases of all channels can be aligned at the LUs, i.e., for $\varepsilon \in \{{\rm R},{\rm T}\}$ we have
\begin{align}
	|{h}_{\varepsilon}|^2 = \left( \sum_{n = 1}^N |h_{r_1,n}||h_{r_2, n}| \right)^2  .
\end{align}

Different from the LUs, phases of different channels are random and independent at the Eves. The overall small-scale fading power is
\begin{align}
	|{h}_{e}|^2 = \left( \sum_{n = 1}^N |h_{BR,n}||h_{e, n}| e^{j \theta_n} \right)^2,
\end{align}
where $\theta_n$ is uniformly distributed in $[0, 2\pi)$.
In this work, the small-scale fading is characterized by the $\kappa$-$\mu$ distribution \cite{doublefading}, which is a general model including some classical distributions such as the Rayleigh, Nakagami-$m$, and Rician as special cases. 
The transmission from the BS to the LUs through the STAR-RIS element $n$ is the double $\kappa$-$\mu$ distribution. The probability density function of the BS-RIS link and the RIS-LU/Eve link is respectively given by
\begin{align}
	f_{|{h}_{r_1,n}|} (x) = \frac{2 \mu_1 (1+ \kappa_1)^{\frac{\mu_1 +1}{2}} x^{\mu_1} e^{- \mu_1 (1+ \kappa_1) x^2}}{{\kappa_1} ^{\frac{\mu_1 -1}{2}} e^{\mu_1 \kappa_1}} \nonumber \\
	\times I_{\mu_1 -1} \left( 2\mu_1 \sqrt{\kappa_1 (1+ \kappa_1)}x \right),
\end{align}
\begin{align}
	f_{|{h}_{r_2,n}|} (x) = \frac{2 \mu_2 (1+ \kappa_2)^{\frac{\mu_2 +1}{2}} x^{\mu_2} e^{- \mu_2 (1+ \kappa_2) x^2}}{{\kappa_2} ^{\frac{\mu_2 -1}{2}} e^{\mu_2 \kappa_2}} \nonumber \\
	\times I_{\mu_2 -1} \left( 2\mu_2 \sqrt{\kappa_2 (1+ \kappa_2)}x \right),
\end{align}
where $I_0(\cdot)$ is the modified Bessel function of the first kind with order zero.

\subsection{STAR-RIS Operation Protocol}
This work considers the TS protocol and the ES protocol to operate the STAR-RIS. Here, We introduce these operation protocols. Since the received signal is highly relative to the change of phases and amplitudes by the STAR-RIS, we provide the transmission- and reflection-coefficient matrixes in two protocols.

\subsubsection{Time Switching Protocol}
In TS protocol, the STAR-RIS operates in the reflecting mode or transmitting mode in different time periods. Let $T_\varepsilon$ denote the percentage of communication time allocated to the LU $\varepsilon$, where $T_{\rm R}+T_{\rm T} =1$ and $T_\varepsilon \in [0,1]$. In the reflecting mode, the transmission- and reflection-coefficient matrix is ${\bf \Theta} _{\rm R}^{\rm TS} = {\tilde {\bf \Theta}}_{\rm R}$ while in the transmitting mode, we have ${\bf \Theta}_{\rm T}^{\rm TS} = {\tilde {\bf \Theta}} _{\rm T}$.

\subsubsection{Energy Splitting Protocol}
In ES protocol, the energy of the signal incident on each element is split into two parts for transmitting and reflecting with energy splitting ratios $\beta_{\rm T}$ and $\beta_{\rm R}$, respectively, and we have $\beta_{\rm R} + \beta_{\rm T} = 1$ according to the law of energy conservation. We consider the same $\beta_\varepsilon$ on all elements of the STAR-RIS. Therefore, the transmission- and reflection-coefficient matrix for LU $\varepsilon$ is ${\bf \Theta}_\varepsilon^{\rm ES} = \sqrt{\beta_\varepsilon} {\tilde {\bf \Theta}}_\varepsilon$.

In this work, we call $T_\varepsilon$ and $\beta_\varepsilon$ the mode operation parameters of TS and ES, respectively. 

\vspace{-0.3 cm}
\subsection{Signal Model}
We focus on the typical LU pair in this work. The remaining LUs use resources that are orthogonal to the typical LU pair, so only intra-cluster NOMA interference occurs in the considered system. In STAR-RIS-NOMA, the SIC process is employed as in traditional NOMA systems. Without loss of generality, the SIC is employed at the LU with the better channel condition in the typical NOMA LU pair to achieve high rate performance. Let $U_s$ and $U_w$ denote the strong LU and the weak LU in the LU pair, respectively. The power allocation coefficient for $U_s$ is $a_s$ and that for $U_w$ is $a_w$, where $a_s +a_w = 1$. For user fairness, the higher power level is always allocated to $U_w$, i.e., $a_w > a_s$.

We consider the BS transmits Gaussian signals for both LUs. If the reflecting LU is the strong LU, i.e., $U_{\rm R} = U_s$, $U_{\rm R}$ decodes the massage of $U_{\rm T}$ first. For the operation protocol ${\rm XS} \in \{{\rm TS}, {\rm ES}\}$, the signal-to-interference-plus-noise ratio (SINR) of the SIC process is given by
\begin{align}
	\gamma_{\rm SIC} ^{\rm XS} = \frac{ c_{\rm R} ^{\rm XS} a_w \rho_b L_{\rm R} \left| h_{\rm R} \right|^2}{ c_{\rm R} ^{\rm XS} a_s \rho_b L_{\rm R} \left|h_{\rm R} \right|^2 + 1} ,
\end{align}
where $c_{\varepsilon} ^{\rm TS} = 1$ and $c_{\varepsilon} ^{\rm ES} = \beta_\varepsilon$ for $\varepsilon \in \{ {\rm T}, {\rm R} \}$. $\rho_b$ is the transmit SNR for LUs. 

After the successful SIC, $U_{\rm R}$ removes the messages of $U_{\rm T}$. Then $U_{\rm R}$ decodes its required messages with the following SNR
\begin{align}
	\gamma_{\rm R} ^{\rm XS} =  c_{\rm R} ^{\rm XS} a_s \rho_b L_{\rm R} \left| h_{\rm R} \right|^2 .
\end{align}

Since $U_{\rm T}$ decodes its message by treating the message of $U_{\rm R}$ as interference, the decoding SINR at $U_{\rm T}$ is expressed as
\begin{align}
	\gamma_{\rm T} ^{\rm XS} = \frac{ c_{\rm T} ^{\rm XS} a_w \rho_b L_{\rm T} \left| h_{\rm T} \right|^2}{ c_{\rm T} ^{\rm XS} a_s \rho_b L_{\rm T} \left| h_{\rm T} \right|^2 + 1} .
\end{align}

For the case that the transmitting LU is the strong LU, the expressions can be obtained similarly, and we skip it here. Since the SIC order in NOMA depends on the order of channel gains, we focus on the performance of the strong LU and the weak LU in the rest of the paper. Thus the subscript $\varepsilon \in\{s, w\}$ related to the type of LUs. 

We consider the worst-case of the secure transmission, and hence we focus on the most detrimental Eve which has the highest detecting SNR of $U_\varepsilon$.
When the most detrimental Eve is at the LU $\tau \in \{s, w\}$ side of the STAR-RIS, the instantaneous SNR of detecting the information of $U_\varepsilon$ at the Eve can be presented as
\begin{align}
	\gamma_{E_\varepsilon} ^{\rm XS} = c_{\tau} ^{\rm XS} a_\varepsilon \rho_e  \mathop{\max}_{i \in {\bf \Phi}_e} \left\{ L_{e,i}\left| h_{e} \right|^2 \right\},
\end{align}
where $\rho_e$ is the transmit SNR for the Eve.

\begin{table*}[t!]
	\centering
	\caption{Typical Fading Models in Cascaded Channel} \label{table: performance}	
	\begin{tabular}{
			|m{1.7 cm}<{\centering} |m{3 cm}<{\centering}
			|m{8 cm}<{\centering}| }
		\hline
		Model & Channel parameters & PDF expression $f_{|h_u|^2} (x)$  \\ \hline 
		double Rayleigh & $\kappa_1 \to 0$, $\kappa_2 \to 0$, $\mu_1 = 1$, $\mu_2 = 1$ & $4x {\rm K}_0 \left( 2x \right) $ \\ \hline
		double Nakagami & $\kappa_1 \to 0$, $\kappa_2 \to 0$, $\mu_1 = m_1$, $\mu_2 = m_2$ & $\frac{4x^{m_1 + m_2 -1}}{\Gamma(m_1) \Gamma(m_2)} (m_1 m_2)^{\frac{m_1 + m_2}{2}} {\rm K}_{m_1 - m_2} \left( 2x \sqrt{m_1m_2} \right)$ \\ \hline
		double Rician & $\kappa_1 = K_1$, $\kappa_2 = K_2$, $\mu_1 = 1$, $\mu_2 = 1$ & $\sum \limits_{r=0}^{\infty} \sum\limits_{t=0}^{\infty} \frac{4x^{r + t +1} K_1^r K_2^t (\Delta_1 \Delta_2)^{\frac{r+t}{2} + 1}}{e^{K_1 + K_2} (r!)^2 (t!)^2} {\rm K}_{r - t} \left( 2x \sqrt{\Delta_1 \Delta_2} \right)$ \\ \hline
	\end{tabular}
\end{table*}

\section{Secrecy Outage Probability}
In this section, we consider the scenario where the CSI of Eves is unavailable at the BS. In this case, we employ the SOP as the performance metric. We first obtain new channel statistics for STAR-RIS-aided links. Then we derive the theoretical SOP expressions of the typical NOMA LU pair in the considered networks. Finally, the asymptotic SOP in the high SNR regime is provided.

\vspace{-0.3 cm}
\subsection{New Channel Statistics}
The STAR-RIS assisted transmission introduces cascaded small-scale fading. For the fading channel from the BS to the LU/Eve through the STAR-RIS element $n$, we denote $\Delta_{n} = |h_{r_1,n}||h_{r_2, n}|$. The probability density function (PDF) of $\Delta_{n}$ can be expressed as \cite{doublefading}
\begin{align}\label{eq: doublefadn pdf}
	f_{\Delta_n} (x) = &\frac{2 \phi_1 \phi_2 x}{e^{\mu_1 \kappa_1 + \mu_2 \kappa_2}} \sum_{q=0}^{\infty} \sum_{t=0}^{\infty} \rho_{q,t} \nonumber \\
	&\times  G_{0,2}^{2,0} \left( \phi_1 \phi_2x^2 \left| q+\mu_1-1, t+\mu_2-1 \right. \right),
\end{align}
where $\phi_i = \mu_i (\kappa_i+1)$ and $\rho_{q,t} = \frac{(\mu_1 \kappa_1)^q (\mu_2 \kappa_2)^t}{q! t! \Gamma(q+\mu_1) \Gamma(t+\mu_2)}$. The $k$-th order moment of the product $\Delta_n$ is given by
\begin{align}
	\mathbb{E} [ (\Delta_n)^k ] = &\frac{(\mu_1)_{\frac{k}{2}} (\mu_2)_{\frac{k}{2}}}{e^{\mu_1 \kappa_1 + \mu_2 \kappa_2} {\phi_1} ^{\frac{k}{2}} {\phi_2} ^{\frac{k}{2}}} {}_1F_1 \left(\frac{k}{2}+\mu_1; \mu_1; \kappa_1\mu_1 \right) \nonumber \\
	&\times  {}_1F_1 \left(\frac{k}{2}+\mu_2; \mu_2; \kappa_2\mu_2 \right),
\end{align}
where $(x)_m = \frac{\Gamma(x+m)}{\Gamma(x)}$ is the pochhammer symbol.

Some widely known fading distributions including Rayleigh distribution, Nakagami-$m$ distribution, and Rician distribution are the special cases of the $\kappa$-$\mu$ distribution. In Table \ref{table: performance}, we conclude the parameters and the simplified PDFs for those typical fading models in the cascaded channel. Allowing $\kappa_i = K_i$ and $\mu_i = 1$ for $i \in\{1,2\}$, we are able to obtain the double Rician distribution with the shape parameter $K_i$ and the scale parameter $1$. When $K_i \to 0$, the distribution becomes the double Rayleigh distribution. If we set $\kappa_i \to 0$ and $\mu_i = m_i$, the double Nakagami-$m$ distribution with the shape parameter $m_i$ and the spread parameter 1.

\begin{lemma}\label{lemma: statistic1}
When the number of STAR-RIS elements is large enough, the overall small-scale fading power for the LUs obeys a Gamma distribution
\begin{align}\label{eq: sfading user}
	|{h}_{u}|^2 \sim {\rm Gamma} \left( \frac{({m _r}^2N +  {\sigma _r}^2)^2 }{\Omega_r},  \frac{\Omega_r N}{{m _r}^2N +  {\sigma _r}^2} \right),
\end{align}
where $\Omega_r = 4{m _r}^2{\sigma _r}^2N + 2{\sigma _r}^4$, $m _r = \mathbb{E} [ \Delta_n]$, and ${\sigma _r}^2 = \mathbb{E} [ (\Delta_n)^2] - \mathbb{E} [ \Delta_n] ^2$. The overall small-scale fading power for the Eves obeys
\begin{align}
	|{h}_{e}|^2 \sim {\rm Gamma} \left( 1, W_e \right),
\end{align}
where $W_e =  N({m_r}^2+{\sigma_r}^2)$.
\end{lemma}

\begin{IEEEproof}
Based on the results in our previous work \cite{STARchannel}, if $m_r$ and ${\sigma_r}^2$ are the mean and the variance of $\Delta_{n}$, respectively, the overall small-scale fading power $|{h}_{u}|^2$ can be approximately fitted by a Gamma distribution ${\rm Gamma} \left( \frac{{M_u}^2}{V_u}, \frac{V_u}{M_u} \right)$, where $M_u = {m _r}^2N^2 +  {\sigma _r}^2N$ and $V_r = 4{m _r}^2{\sigma _r}^2N^3 + 2{\sigma _r}^4N^2$. Moreover, the overall small-scale fading power for the Eves fulfills $|{h}_{e}|^2 \sim N({m_r}^2+{\sigma_r}^2) \times {\rm Gamma} \left( 1, 1 \right)$. According to the property of the cascaded $\kappa$-$\mu$ distribution, this lemma is proved.
\end{IEEEproof}

\begin{lemma}\label{lemma: odered cp CDF}
In the NOMA LU pair, CDFs of the channel power for the strong LU and the weak LU can be respectively expressed as
\begin{align}
	F_{H_{s}}(x) &= [ \hat{F}_{H_u}(x) ]^2 ,\\
	F_{H_{w}}(x) &= 2\hat{F}_{H_u}(x) - [ \hat{F}_{H_u}(x) ]^2,
\end{align}
where $\hat{F}_{H_u}(x) = \frac{\delta}{\Gamma (k_r)} G_{2,3}^{1,2} \left( \frac{{R_U}^\alpha x}{A_L \theta_r}  \left| \begin{matrix}
	1-\delta,1 \\ k_r,0,-\delta
\end{matrix} \right. \right) $, $\delta = \frac{2}{\alpha}$, $A_L = C_r {l_{BR}}^{-\alpha}$, $k_r = \frac{({m _r}^2N +  {\sigma _r}^2)^2 }{4{m _r}^2{\sigma _r}^2N + 2{\sigma _r}^4}$, and $\theta_r = \frac{4{m _r}^2{\sigma _r}^2N^2 + 2{\sigma _r}^4N}{{m _r}^2N +  {\sigma _r}^2}$.
\end{lemma}
\begin{IEEEproof}
See Appendix~A.
\end{IEEEproof}

\subsection{Secrecy Outage Probability Analysis}
For the protocol ${\rm XS} \in \{{\rm TS}, {\rm ES}\}$, let $C_{U_\varepsilon} ^{\rm XS}$ denote the channel capacity of the pair of LUs and $C_{E_\varepsilon} ^{\rm XS}$ represent the channel capacity of the most detrimental Eve with the data of $U_\varepsilon$. Then the secrecy capacity of the NOMA LUs can be expressed as $\left[ C_{U_\varepsilon} ^{\rm XS} - C_{ E_\varepsilon} ^{\rm XS} \right]^+$
which is non-negative. For a target rate $R_\varepsilon$, if $C_\varepsilon ^{\rm XS} \ge R_\varepsilon$, the information can be transmitted to $U_\varepsilon$ in perfect secrecy. Otherwise, the information-theoretic security is compromised. Therefore, the SOP of $U_\varepsilon$ can be defined as
\begin{align}\label{eq: defsop oma}
	P_\varepsilon ^{\rm XS} = {\rm Pr} \left( \left[ C_{U _\varepsilon} ^{\rm XS} - C _{E _\varepsilon} ^{\rm XS} \right]^+ < R_\varepsilon \right).
\end{align}

We can observe from \eqref{eq: defsop oma} that to calculate the SOP, it is important to derive the probability distribution of $C_{U _\varepsilon} ^{\rm XS}$ and $C _{E _\varepsilon} ^{\rm XS}$. In the following, we first provide the statistics of $C_{U _\varepsilon} ^{\rm XS}$ and $C _{E _\varepsilon} ^{\rm XS}$. On this basis, the SOP expressions are derived.

Note that the most detrimental Eve is at either the strong LU side or the weak LU side, we provide the channel statistics of the most detrimental Eve at the LU $\tau \in \{s,w\}$ side.

\begin{lemma}\label{lem: Eve SNR oneside}
For the protocol ${\rm XS} \in \{{\rm TS}, {\rm ES}\}$, the CDF of the received SNR $\gamma_{E_{\varepsilon, \tau}}^{\rm XS}$ at the most detrimental Eve $E_\tau$ $(\tau \in \{s,w\})$ in terms of the message of $U_\varepsilon$ is given by
\begin{align}\label{eq: Eve cg}
	F_{\gamma_{E_{\varepsilon, \tau}}^{\rm XS}}(x) =  \exp \left({- m_\varepsilon \left( \frac{x}{c_{\tau} ^{\rm XS} }\right)^{-\delta}} \right),
\end{align}
where $m _\varepsilon = \frac{1}{2}\pi \delta \lambda_e(\rho_e a_\varepsilon A_L W_e)^{\delta} \Gamma(\delta)$ and $\delta = \frac{2}{\alpha}$.
\end{lemma}

\begin{IEEEproof}
The CDF of the channel gain for the most detrimental Eve can be calculated as follows
\begin{align}\label{eq: Eve cg p1}
	F_{\gamma_{E_{\varepsilon, \tau}}^{\rm XS}}(x) = \mathbb{E}_{{\bf \Phi}_e}\left[\prod _{{\bf \Phi}_e} F_{|h_e|^2} \left(\frac{{d_e}^{\alpha} x}{\rho_e a_\varepsilon A_L c_{\tau}^{\rm XS}} \right) \right].
\end{align}

We apply the probability generating functional \cite[eq. (4.3)]{GPP} and utilizing the property that ${\rm Gamma}(1,W_e) = W_e {\rm Exp}(1)$. The \eqref{eq: Eve cg p1} can be rewitten as
\begin{align}
	F_{\gamma_{E_{\varepsilon, \tau}}^{\rm XS}}(x) &= \exp \bigg(- \pi \lambda_e \int_{0}^{\infty} r \nonumber \\ &\times \bigg(1- F_{|h_e|^2} \left(\frac{{r}^{\alpha} x}{\rho_e a_\varepsilon A_L c _{\tau} ^{\rm XS} W_e} \right) \bigg)  dr  \bigg) \nonumber \\
	&\overset{(a)}{=} \exp \left(- \frac{ \pi \lambda_e(\rho_e a_\varepsilon A_L c_{\tau} ^{\rm XS}W_e )^{\delta}\Gamma( \delta)}{\alpha x^{\delta}} \right),
\end{align}
where $(a)$ is obtained by applying \cite[eq. (3.326.10)]{Intetable}. This completes the proof.
\end{IEEEproof}

When adopting the TS protocol, the channel capacity for the pair of LUs is $C_{U _\varepsilon} ^{\rm TS} = T _\varepsilon \log_2(1+\gamma _\varepsilon ^{\rm TS})$, while the capacity for the reflecting/transmitting Eve is $C_{E_{\varepsilon, \tau}} ^{\rm TS} = T_\tau \log_2(1+\gamma_{E_{\varepsilon, \tau} } ^{\rm TS})$. To obtain neat expressions, we propose a mapping method to unify the SNR at different sides of the STAR-RIS. We introduce the equivalent received SNR for Eves as follows.
\begin{definition}\label{def: equiv SNR TS}
For the TS protocol, we deploy an equivalent Eve $E_{\tau \to \varepsilon}$ of the most detrimental $E_\tau$, which locates at the same side of the STAR-RIS as the $U_\varepsilon$ and has the same capacity as $E_\tau$. The equivalent received SNR at $E_{\tau \to \varepsilon}$ is
\begin{align}
	{\hat \gamma}_{E_{\tau \to \varepsilon}}^{\rm TS} = \left( 1 + \gamma_{E_\tau} ^{\rm TS} \right)^{ \frac{T_\tau}{T_\varepsilon}} -1.
\end{align}
\end{definition}

Then the PDF of the equivalent received SNR of the most detrimental Eve in all reflecting and transmitting Eves can be calculated.
\begin{lemma}
For the TS protocol, the PDF of the equivalent received SNR for the most detrimental Eve in terms of the $U _\varepsilon$ data is given by
\begin{align}
	f_{{\hat \gamma}_{E _{\varepsilon}} ^{\rm TS}} (x) =  \delta m _\varepsilon e^{- m_e \sum _{\tau \in \{s,w\}}  \left( (x+1)^{{\tilde T}_{\varepsilon,\tau}} -1\right)^{-\delta} } \nonumber \\  \times \sum _{\tau \in \{s,w\}}  {\tilde T}_{\varepsilon,\tau} (x+1)^{{\tilde T}_{\varepsilon,\tau} -1} \nonumber 
	\\ \times \left( (x+1)^{{\tilde T}_{\varepsilon,\tau} }  -1\right)^{-\delta-1},
\end{align}
where ${\tilde T}_{\varepsilon,\tau} = \frac{T _\varepsilon}{T _\tau}$.
\end{lemma} 

\begin{IEEEproof}
Based on {\bf Definition~\ref{def: equiv SNR TS}} and the results in {\bf Lemma~\ref{lem: Eve SNR oneside}}, the CDF of the equivalent received SNR for the Eve is expressed as
\begin{align}
	F_{{\hat \gamma}_{E_\varepsilon}^{\rm TS}}(x) &= F_{\gamma_{E_{\varepsilon,{\rm T}}}^{\rm TS}}((x+1)^\frac{T _\varepsilon}{T_{\rm T}} -1)  F_{\gamma_{E_{\varepsilon,{\rm R}}}^{\rm TS}}((x+1)^\frac{T _\varepsilon}{T_{\rm R}} -1) \nonumber \\ 
	&=  e^{- m_\varepsilon \bigg( \left( (x+1)^\frac{T _\varepsilon}{T_{\rm T}} -1\right)^{-\delta} + \left( (x+1)^\frac{T _\varepsilon}{T_{\rm R}} -1\right)^{-\delta} \bigg) }.
\end{align}
By taking the derivative of $F_{{\hat \gamma}_{E _\varepsilon}^{\rm TS}}(x)$, we obtain the PDF of ${\hat \gamma}_{E _\varepsilon}^{\rm TS}$. Note that $\tau \in \{{\rm R}, {\rm T}\}$ is equivelent to $\tau \in \{s, w\}$ in the expression of $F_{{\hat \gamma}_{E_\varepsilon}^{\rm TS}}(x)$, the lemma is proved.
\end{IEEEproof}

\begin{theorem}\label{theo: NOMA TS}
For the TS protocol, the SOPs of the two NOMA LUs are given by
\begin{align}\label{eq: SOP TS s}
	P_s^{\rm TS} =  \int_0^\infty F_{H_s} \left( \frac{2^{\frac{R_s}{T_s}} (x+1) -1}{a_s \rho_b} \right) f_{{\hat \gamma}_{E_s}^{\rm TS}}(x) dx,
\end{align}
\begin{align}
	P_w^{\rm TS} = \int_0^{B_{up}^{\rm TS}} F_{H_w} \left( \frac{1}{\rho_b} \frac{2^{\frac{R_w}{T_w}} (x+1) -1}{a_w - a_s \left( 2^{\frac{R_w}{T_w}} (x+1) -1 \right)} \right) \nonumber \\
	\times  f_{{\hat \gamma}_{E_w}^{\rm TS}}(x) dx 
	+ {\bar F}_{{\hat \gamma}_{E_w}^{\rm TS}}\left(B_{up}^{\rm TS}\right),
\end{align}
where $B_{up}^{\rm TS} = \frac{1}{2^{R_w/T_w} a_s} - 1$. 
\end{theorem}

\begin{IEEEproof}
For both LUs, the SOP is related to $f_{{\hat \gamma}_{E_{\varepsilon}} ^{\rm TS}}(x)$ and the CDF of received SNR of the LU. The SOP for the strong LU can be expressed as
\begin{align}
	P_{s} ^{\rm TS} = \int_0^\infty f_{{\hat \gamma}_{E_s}^{\rm TS}}(x) F_{\gamma_{U_s} ^{\rm TS}} \left( 2^{R_s/T_s} ( x +1) -1 \right) dx.
\end{align}
Based on the fact that $\gamma_{U_s} ^{\rm TS} = a_s \rho_b H_s$,  \eqref{eq: SOP TS s} is obtained. For the weak LU, since the outage probability is 1, i.e., $F_{\gamma_{U_s} ^{\rm TS}}(x) = 1$ when $a_w - \left(2^{\frac{x}{T_w}} ({\hat \gamma}_{E_s}^{\rm TS}+1) -1\right)a_s \le 0$, the SOP for the weak LU consists of two parts
\begin{align}
	P_{w} ^{\rm TS} = \int_0^{B_{up}^{\rm TS}} f_{{\hat \gamma}_{E_w}^{\rm TS}}(x) F_{\gamma_{U_w} ^{\rm TS}} \left( 2^{R_w/T_w} ( x +1) -1 \right) dx \nonumber \\ + \int_{B_{up}^{\rm TS}}^\infty f_{{\hat \gamma}_{E_w}^{\rm TS}}(x)dx.
\end{align}
Utilizing $\gamma_{U_w} ^{\rm TS} = \frac{a_w \rho_b H_w}{a_s \rho_b H_w + 1}$, the proof is completed.
\end{IEEEproof}

From {\bf Theorem \ref{theo: NOMA TS}}, we can easily observe that with the increase of $T_s$, the secrecy outage performance for the strong LU is improved monotonically while the trend is the opposite for the weak LU. Therefore, there is a trade-off between the SOPs of the strong LU and the weak LU.

When considering the ES protocol, the channel capacity of the legitimate LUs is expressed as $C_{U _\varepsilon} ^{\rm ES} = \log_2 (1+ \gamma_\varepsilon ^{\rm ES})$ and that for the Eves is $C_{E _\varepsilon} ^{\rm ES} =\log_2 (1+ \gamma_{E _\varepsilon}^{\rm ES}) $. Similarly, we utilize the mapping method for the equivalent received SNR in this case as follows.

\begin{definition}\label{def: equiv SNR ES}
For the ES protocol, we deploy an equivalent Eve of $E_\tau$ located at the same side as the $U_\varepsilon$, and the equivalent received SNR is
\begin{align}
	{\hat \gamma}_{E_{\tau \to \varepsilon}} ^{\rm ES} = \frac{\beta_\tau}{\beta_\varepsilon} \gamma_{E_\varepsilon} ^{\rm ES} .
\end{align}
\end{definition}

\begin{lemma}
For the ES protocol, the PDF of the equivalent received SNR at the $U_\varepsilon$ side for the most detrimental Eve is given by
\begin{align}
	f_{{\hat \gamma}_{E_{\varepsilon}} ^{\rm ES}} (x) = \delta m_\varepsilon e^{-   m_\varepsilon \sum_{\tau \in \{s,w\}} \left( {\tilde \beta}_{\varepsilon, \tau} x\right)^{-\delta}} \sum_{\tau \in \{s,w\}} {{\tilde \beta}_{\varepsilon, \tau}} ^{-\delta} x^{-\delta-1} ,
\end{align}	
where ${\tilde \beta}_{\varepsilon, \tau} = \frac{\beta \varepsilon}{{\beta _\tau}}$.
\end{lemma} 

\begin{IEEEproof}
We have the CDF of ${\hat \gamma}_{E_{\varepsilon}} ^{\rm ES}$ expressed as $F_{{\hat \gamma}_{E_{\varepsilon}} ^{\rm ES}} (x) = e^{-   m_\varepsilon \left( {\tilde \beta}_{\varepsilon, s} x\right)^{-\delta} - m_\varepsilon \left( {\tilde \beta}_{\varepsilon, w} x\right)^{-\delta}}$ based on {\bf Lemma~\ref{lem: Eve SNR oneside}}, then this lemma is straightforwardly proved.
\end{IEEEproof}

\begin{theorem}\label{theo: NOMA ES}
For the ES protocol, the SOPs of the two NOMA LUs are given by
\begin{align}
	P_s^{\rm ES} = \int_0^\infty  &F_{H_s} \left( \frac{2^{R_s} ( \beta_s x+1) -1}{\beta_s a_s \rho_b} \right)  f_{{\hat \gamma}_{E_{s}} ^{\rm ES}} (x) dx,
\end{align}
\begin{align}
	P_w^{\rm ES} = \int_0^{B_{up}^{\rm ES}}  F_{H_w} \left( \frac{1}{\rho_b \beta_w } \frac{2^{R_w} (\beta_wx+1) -1}{a_w - a_s \left( 2^{R_w} (\beta_wx+1) -1 \right)} \right)  \nonumber \\
	\times  f_{{\hat \gamma}_{E_{w}} ^{\rm ES}} (x) + {\bar F}_{{\hat \gamma}_{E_{w}} ^{\rm ES}} (B_{up}^{\rm ES}),
\end{align}
where $B_{up}^{\rm ES} = \frac{1}{2^{R_w} a_s \beta_w} - \frac{1}{\beta_w}$.
\end{theorem}

\begin{IEEEproof}
The proof is similar to {\bf Theorem \ref{theo: NOMA TS}} and hence we skip it here.
\end{IEEEproof}

\begin{corollary}
For the protocol ${\rm XS} \in \{{\rm TS}, {\rm ES}\}$, the SOPs of two LUs in the NOMA LU pair have the approximated closed-form expressions
\begin{align}
	P_s^{\rm XS} \approx \sum_{m = 1}^{M_s} \frac{\xi_m G_s^{\rm XS} \left( \xi_m  \right)}{(M_s + 1)^2 [L_{M_s + 1}(\xi_m)]^2  \exp(-\xi_m)}  ,
\end{align}
\begin{align}
	P_w^{\rm XS} \approx \sum_{m = 1}^{M_w} \frac{\pi B_{up}^{\rm XS}}{2M_w} \sqrt{1 - {\varphi_m}^2} G_w^{\rm XS} \left( \frac{B_{up}^{\rm XS}}{2}\varphi_m + \frac{B_{up}^{\rm XS}}{2} \right) \nonumber \\
	+ {\bar F}_{{\hat \gamma}_{E_{w}} ^{\rm XS}} (B_{up}^{\rm XS}) ,
\end{align}
where $G_s ^{\rm XS}(x) = f_{{\hat \gamma}_{E_{s} ^{\rm XS}}}(x) F_{H_s} \left( \frac{1}{\rho_b} g_s ^{\rm XS}(x) \right)$, $G_w(x) = f_{{\hat \gamma}_{E_{w} ^{\rm XS}}}(x) F_{H_w} \left( \frac{1}{\rho_b} g_w ^{\rm XS}(x) \right)$, $g_s ^{\rm TS}(x) = 2^{\frac{R_s}{T_s}} (x+1)-1$, $g_s ^{\rm ES}(x) = \frac{2^{R_s} (\beta_s x+1)-1}{\beta_s}$, $g_w ^{\rm TS}(x) = \frac{2^{\frac{R_w}{T_w}} (x+1)-1}{a_w - a_s \left( 2^{\frac{R_w}{T_w}} (x+1)-1 \right)}$, and $g_w ^{\rm ES}(x) = \frac{2^{R_s} (\beta_s x+1)-1}{a_w - a_s \left( 2^{R_s} (\beta_s x+1)-1 \right)}$. 
$\varphi_m = \cos\left(\frac{2m-1}{2M_w} \pi \right)$ and $\xi_m$ is the $m$-th root of Laguerre polynomial $L_{M_s} (x)$. $M_s$ and $M_w$ are parameters to ensure a complexity-accuracy trade-off. 
\end{corollary}
\begin{IEEEproof}
By applying the Gauss–Laguerre quadrature and the Chebyshev–Gauss quadrature to the SOP expressions for the strong LU and the weak LU, respectively, the corollary can be proved.
\end{IEEEproof}

Similar to the conclusion of the TS protocol, with a larger $\beta_s$ the better SOP for the strong LU is obtained while the secrecy performance becomes worse for the weak LU in the ES protocol. 

\begin{remark}
When adjusting the STAR-RIS mode operation parameters $T_\varepsilon$ or $\beta_\varepsilon$, there is a trade-off between the SOP performance of the strong LU and the weak LU. In practice implementation, we should decide the STAR-RIS mode operation parameters according to different LU requirements.
\end{remark}


\subsection{Diversity Order Analysis}
To show further insights into the system implementation, we investigate the SOP in the high-SNR regime. Eves are assumed to have a powerful detection capability as the above analysis. The asymptotic performance is analyzed, especially when the difference of the channel SNR between the BS and LUs is sufficiently high, i.e., $\rho_b \to \infty$. Note that when Eve’s transmit SNR $\rho_e \to \infty$, the probability of successful eavesdropping tends to unity. We define the secrecy diversity order as
\begin{align}
	{\cal D} ^{\rm XS} = - \lim \limits_{\rho_b \to \infty} \frac{ \log_2 \left(P_{out, \infty} ^{\rm XS} \right) }{\log_2 (\rho_b)},
\end{align}
where $P_{out, \infty} ^{\rm XS}$ is the asymptotic SOP.

It can be observed from {\bf Theorem \ref{theo: NOMA TS}} and {\bf Theorem \ref{theo: NOMA ES}} that the SOP is the integral of the product of $f_{{\hat \gamma}_{E _{\varepsilon} }^{\rm XS}}(x)$ and $F_{H_\varepsilon} (x)$. The expression is quite complex, so the relationship between $\rho_b$ and the SOP expression is not straightforward. We first derive the asymptotic CDF of the unordered channel power gain $\hat{F} _{H _u}(x)$ which is related to $F_{H_\varepsilon} (x)$.

\begin{lemma}\label{lemma: exact order}
When $x \to 0^+$, the CDF of unordered LU channel power gain is given by
\begin{align}
	\hat{F}_{H_{u}}^{0^+}(x) = L_u x^{\hat{\mu}N},
\end{align}
where $L_u = \frac{2 {A_u}^N {R_U}^{\alpha\hat{\mu}N}}{ {A_L}^{\hat{\mu}N}}$, $\hat{\mu} = \min \{\mu_1, \mu_2\}$ and  $A_u = \frac{ K_u \rho_{0,0} \left(\phi_1 \phi_2\right)^{\hat{\mu}} \Gamma(|\mu_1-\mu_2|) \Gamma(\frac{1}{2}+\hat{\mu}) \Gamma(\hat{\mu}) }{\sqrt{\pi}e^{\mu_1 \kappa_1 + \mu_2 \kappa_2}}$. $K_u = 2$ when $\mu_1 = \mu_2 $; otherwise, $K_u = 1$.
\end{lemma}

\begin{IEEEproof}
	See Appendix~B.
\end{IEEEproof}

Based on \eqref{eq: A.5} in Appendix A, the asymptotic CDF of ordered LU channel power gain is $F_l^{0^+}(x) = \sum_{k = l}^{K} \binom{K}{k} [L_u x^{\hat{\mu}N}]^{k} [1-L_u x^{\hat{\mu}N}]^{K-k} \approx {L_u}^l x^{\hat{\mu}Nl}$ for the $l$-th weakest LU. If the signal is transmitted to LUs without being eavesdropped, i.e., $\rho_e \to 0$, the outage probability can be expressed as $F_{H _\varepsilon} \left(\frac{1}{\rho_b} g_\varepsilon ^{\rm XS}(0) \right)$. Therefore, the diversity orders for the weak LU and the strong LU are $\hat{\mu}N$ and $2 \hat{\mu}N$, respectively.

\begin{remark}
When considering the STAR-RIS-assisted transmission in the no-eavesdropping scenario, the diversity order for the NOMA LUs has a linear correlation with the number of elements on the STAR-RIS. Therefore, the outage performance can be improved by increasing the number of elements in this case.
\end{remark}

When considering the PLS, however, we cannot ignore the impact of $f_{{\hat \gamma}_{E _{\varepsilon} } ^{\rm XS}}(x)$ on the SOP performance. Since ${\hat \gamma}_{E _{\varepsilon} } ^{\rm XS}$ may have a long-tail PDF, it is unreasonable to calculate the secrecy diversity order by utilizing the asymptotic CDF obtained in {\bf Lemma \ref{lemma: exact order}} as in \cite{LiuSOP}. Therefore, we employ the method of changing variables to calculate the secrecy diversity order of the strong LU as follows.

\begin{corollary}\label{collo: aSOP su}
In STAR-RIS-NOMA, the secrecy diversity order of the strong LU is expressed as
\begin{align}
	{\cal D}^{\rm XS}_s = \begin{cases} 
		\hat{T} \delta &{\rm XS} = {\rm TS} \\ \delta &{\rm XS} = {\rm ES},  \end{cases} 
\end{align}
where $\hat{T} = \min \limits_{\tau \in \{s,w\}} \hat{T}_{\tau, s}$.
\end{corollary}

\begin{IEEEproof}
Noticed that only $F_{H_s} \left(\frac{1}{\rho_b} g_s^{\rm XS}(x) \right)$ includes $\rho_b$ while the accurate expression of  $F_{H_s} \left(\frac{1}{\rho_b} g_s^{\rm XS}(x) \right)$ is quite complicated, we calculate the SOP by changing the variable $t = \frac{x}{\rho_b}$. For the TS protocol, the SOP for the strong LU is rewritten as
\begin{align}
	P_s^{\rm TS} = \delta m_s \rho_b \int_0^{\infty} F_{H_s} \left(\frac{1}{\rho_b} g_s^{\rm TS}(\rho_b t) \right) \nonumber \\ \times e^{- m_s \sum _{\tau \in \{s,w\}} \left( (\rho_b t+1)^{{\hat T}_{s,\tau}} -1\right)^{-\delta} } \nonumber \\
	\times    \sum _{\tau \in \{s,w\}}  {\hat T}_{s,\tau} (\rho_b t +1)^{{\hat T}_{s,\tau} -1} \nonumber \\ \times \left( (\rho_b t+1)^{{\hat T}_{s,\tau} }  -1\right)^{-\delta-1} dt.
\end{align}
Setting $\rho_b \to \infty$, we have
\begin{align}
	P_{s,\infty}^{\rm TS} \approx &\delta m_s \rho_b \int_0^{\infty} F_{H_s} \left( 2^{\frac{R_s}{T_s}} t \right) \nonumber \\ 
	&\times \sum _{\tau \in \{s,w\}}  {\hat T}_{s,\tau} (\rho_b t + 1)^{ -{\hat T}_{s,\tau} \delta -1}  dt.
\end{align}
Then we are able to obtain the secrecy diversity order by using its definition. For the ES protocol, the derivation procedure is similar and hence we skip it here. The proof is completed.
\end{IEEEproof}

\begin{remark}
Since $\hat{T} \le 1$, the secrecy diversity order of the strong LU in the TS protocol is no higher than that in the ES protocol.
\end{remark}

Different from the scenario without Eves, the secrecy diversity order for the strong LU is unrelated to the number of STAR-RIS elements. Furthermore, by properly adjusting the STAR-RIS mode operation parameter, the best secrecy diversity order for both protocols is $\delta$, which is only determined by the path loss exponent.

For the weak LU, intra-cluster interference may degrade the secrecy performance. As we have discussed in the proof of {\bf Theorem \ref{theo: NOMA TS}}, if 
\begin{align}
	a_w - \left(2^{\frac{R_w}{T_w}} ({\hat \gamma}_{E_s}^{\rm TS}+1) -1\right)a_s \le 0,
\end{align} 
holds, the SOP of the weak LU is one for the TS protocol. In this case, the SOP has an error floor due to $F_{{\hat \gamma}_{E_s}^{\rm TS}}(\frac{1}{2^{R_w/T_w} a_s}) < 1$. A similar conclusion can be obtained for the ES protocol. 
\begin{remark}
Considering the PLS in the NOMA scheme, the error floor of the weak LU in terms of the SOP depends on the CDF of the received SNR for the Eve, i.e., $F_{{\hat \gamma}_{E_s}^{\rm TS}}(x)$. When $F_{{\hat \gamma}_{E_s}^{\rm TS}}(B_{up} ^{\rm XS}) = 1$, the error floor can be avoided. This usually happens when the SNR is high and the required threshold is low.
\end{remark}

Based on the characteristic of $f_{{\hat \gamma}_{E_s}^{\rm TS}}(x)$, we are able to calculate the error floor in the following corollary.
\begin{corollary}\label{collo: aSOP wu}
The secrecy error floor of the weak LU is expressed as
\begin{align}
	{ EF}^{\rm XS}_w = \bar{F}_{\hat{\gamma}_w ^{\rm XS}} \left( B_{up} ^{\rm XS} \right).
\end{align}
Therefore, the secrecy security order of the weak LU is zero.
\end{corollary}
\begin{IEEEproof}
When $\rho_b \to \infty$, the received SINR for the weak LU tends to a constant $\frac{a_w}{a_s}$. Let us take the TS protocol as an example. Based on the definition of the SOP, in the high SNR, we have
\begin{align}
	P_s ^{\rm TS} &= {\rm Pr} \left( C_{E_w} ^{\rm TS} > T_w \log_2 \left( 1 + \frac{a_w}{a_s} \right) - R_w \right) \nonumber \\
	&= {\rm Pr} \left( \log_2 \left( \hat{\gamma}_w ^{\rm XS} +1 \right) >  \log_2 \left( 1 + \frac{a_w}{a_s} \right) - \frac{R_w}{T_w} \right),
\end{align}
then the corollary is proved.
\end{IEEEproof}


\begin{remark}
In the considered secure STAR-RIS-NOMA transmission, the SOP of the NOMA LU pair has the error floor due to the error floor of the weak LU. Therefore, the secrecy diversity order of the NOMA LU pair is zero.
\end{remark}

\section{Average Secrecy Capacity}
In this section, we consider the scenario where the CSI of Eves is available at the BS. Here we employ the ASC as the principal secrecy performance metric because the BS can adapt transmission rate according to CSI of the LUs and the Eves to achieve perfect secure transmission. The closed-form expressions are derived first, and then the asymptotic ASC, i.e., the secrecy slope, is investigated for further insights.

\subsection{Average Secrecy Capacity Analysis}
The ASC is defined as the expectation value of the non-negative secrecy capacity over the fading channel and the spatial effect \cite{ASC}. In the considered networks, the ASC for the protocol ${\rm XS} \in \{{\rm TS}, {\rm ES}\}$ is expressed as
\begin{align}\label{def: ASC}
	C_{\varepsilon} ^{\rm XS} = \mathbb{E} \left[ \left[ C_{U _\varepsilon} ^{\rm XS} - C_{E _\varepsilon} ^{\rm XS}  \right]^+ \right].
\end{align}

We observe that the theoretical expression of ASC can be obtained based on the derivation of the SOP. The analtyical expressions of the ASC for the TS protocol and ES protocol are given in {\bf Theorem \ref{theo: cap TS}} and {\bf Theorem \ref{theo: cap ES}}.
\begin{theorem}\label{theo: cap TS}
	For the TS protocol, the ASC expressions for the two NOMA LUs are given by
	\begin{align}\label{eq: cap TS s}
		C_{s} ^{\rm TS} = \frac{T_s}{\ln2} \int_0^{\infty} \frac{{\bar F} _{\gamma _{U_s}}(x) F _{\gamma _{E_s}}(x) }{1+x} dx ,
	\end{align}
	\begin{align}\label{eq: cap TS w}
		C_{w} ^{\rm TS} = \frac{T_w}{\ln2} \int_0^{\frac{a_w}{a_s}} \frac{{\bar F} _{\gamma _{U_w}}(x) {F} _{\gamma _{E_w}}(x)  }{1+x} dx.
	\end{align}
\end{theorem}
\begin{IEEEproof}
	See Appendix~C.
\end{IEEEproof}

\begin{theorem}\label{theo: cap ES}
	For the ES protocol, the ASC expressions for the two NOMA LUs are as follows
	\begin{align}
		C_{s} ^{\rm ES} = \frac{1}{\ln2} \int_0^{\infty} \frac{{\bar F} _{\gamma _{U_s}}(x / \beta_s) F _{\gamma _{E_s}}(x / \beta_s) }{1+x} dx ,
	\end{align}
	\begin{align}
		C_{w} ^{\rm ES} = \frac{1}{\ln2} \int_0^{\frac{a_w}{a_s}} \frac{{\bar F} _{\gamma _{U_w}}(x / \beta_w) {F} _{\gamma _{E_w}}(x / \beta_w)  }{1+x} dx.
	\end{align}
\end{theorem}
\begin{IEEEproof}
	By utilizing similar proof of {\bf Theorem \ref{theo: cap TS}}, this theorem can be proved.
\end{IEEEproof}
\begin{remark}\label{remark: Linearrate TS}
	For the TS protocol, the ASC is linearly related to the STAR-RIS mode operation parameter $T_\varepsilon$. For the ES protocol, however, the ASC is non-linear with $\beta_s$.
\end{remark}

Afterwards, the derived ASC can be further written as closed-form expressions in the following corollaries.
\begin{corollary}\label{coro: cap TS}
For the TS protocol, the closed-form approximations of the ASC for the two NOMA LUs are expressed as
\begin{align}
	C_{s} ^{\rm TS} \approx \frac{T_s}{\ln2} \sum_{m = 1}^{M_s} \frac{\xi_m {\bar F} _{\gamma _{U_s}} \left( \xi_m  \right) {F} _{\gamma _{E_s}} \left( \xi_m  \right) /(1+\xi_m )}{(M_s + 1)^2 [L_{M_s + 1}(\xi_m)]^2  \exp(-\xi_m)} ,
\end{align}
\begin{align}
	C_{w} ^{\rm TS} \approx &\frac{T_w}{\ln2} \sum_{m = 1}^{M_w}    \frac{\pi a_w \sqrt{1 - {\varphi_m}^2} }{\left(a_w \varphi_m+ a_s +1 \right) M_w} {\bar F} _{\gamma _{U_w}} \left(  \frac{a_w(\varphi_m+1)}{2a_s} \right) \nonumber \\
	&\times   {F} _{\gamma _{E_w}} \left(  \frac{a_w(\varphi_m+1)}{2a_s} \right) .
\end{align}
\end{corollary}
\begin{IEEEproof}
By applying the Gauss-Laguerre quadrature and Chebyshev-Gauss quadrature to \eqref{eq: cap TS s} and \eqref{eq: cap TS w}, respectively, the closed-form approximations can be obtained.
\end{IEEEproof}

\begin{corollary}\label{coro: cap ES}
For the ES protocol, the closed-form approximations of the ASC for the two NOMA LUs are given by
\begin{align}
	C_{s} ^{\rm ES} \approx \frac{1}{\ln2} \sum_{m = 1}^{M_s} \frac{\xi_m {\bar F} _{\gamma _{U_s}} \left( \xi_m /\beta_s \right) {F} _{\gamma _{E_s}} \left( \xi_m/\beta_s  \right) /(1+\xi_m )}{(M_s + 1)^2 [L_{M_s + 1}(\xi_m)]^2  \exp(-\xi_m)} ,
\end{align}
\begin{align}
	C_{w} ^{\rm ES} \approx &\frac{1}{\ln2} \sum_{m = 1}^{M_w} \frac{\pi a_w \sqrt{1 - {\varphi_m}^2} }{\left(a_w \varphi_m+ a_s +1 \right) M_w}  {\bar F} _{\gamma _{U_w}} \left(  \frac{a_w(\varphi_m+1)}{2a_s\beta_w} \right) \nonumber \\
	&\times  {F} _{\gamma _{E_w}} \left(  \frac{a_w(\varphi_m+1)}{2a_s\beta_w} \right).
\end{align}
\end{corollary}
\begin{IEEEproof}
The proof is similar to {\bf Corollary \ref{coro: cap TS}}.
\end{IEEEproof}

\begin{proposition}
The ASC of the typical NOMA LU pair is given by
\begin{align}
	C ^{\rm XS} = C_{s} ^{\rm XS} + C_{w} ^{\rm XS}.
\end{align}
\end{proposition}

\begin{table}[t!]
	\centering
	\caption{Secrecy Diversity Order and Secrecy Slope for Different STAR-RIS Protocols} \label{table: sumasmptotic}	
	\begin{tabular}{
			|m{1.3 cm}<{\centering}|m{0.9 cm}<{\centering} |m{1.5 cm}<{\centering}|m{1.5 cm}<{\centering}|m{1.3 cm}<{\centering}
			| }
		\hline
		\multirow{2}{*}{Protocol} & \multirow{2}{*}{LU} & \multicolumn{2}{c|}{Secrecy Diversity} & \multirow{2}{*}{\makecell[c]{Secrecy\\ Slope}} \\ \cline{3-4}
		&  & $\rho_e >0$ & $\rho_e \to 0$ &  \\ \hline 
		\multirow{2}{*}{TS} &SU & $2 \hat{T} /\alpha$ & $2\hat{\mu}N$ & $T_s$ \\ \cline{2-5}
		&WU & 0 & $\hat{\mu}N$ & 0 \\ \hline
		\multirow{2}{*}{ES}& SU & $2/\alpha$ & $2\hat{\mu}N$& 1 \\ \cline{2-5}
		&WU & 0 & $\hat{\mu}N$& 0 \\ \hline
	\end{tabular}
\end{table}

\begin{figure}[t!]
	\centering
	\includegraphics[width= 3 in]{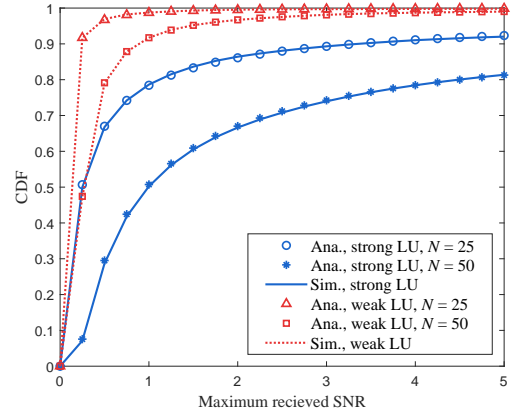}
	\caption{CDF of the maximum received SNR at LUs with $\rho_b = 50$ dB.}\label{fig: channelstatistic}
	\vspace{-0.1 cm}
\end{figure}

\subsection{Secrecy Slope Analysis}
Similarly, to gain insights into the ASC performance, the secrecy slope in the high-SNR regime is considered, which is defined as
\begin{align}
	{\cal S}^{\rm XS} = \lim \limits_{\rho_b \to \infty} \frac{C _{\infty} ^{\rm XS}}{\log_2(\rho_b)},
\end{align}
where $C _{\infty} ^{\rm XS}$ is the asymptotic ASC when $\rho_b \to \infty$. The asymptotic expressions for the pair of NOMA LUs are provided in the following propositions.

\begin{figure*}[t!]
	\normalsize
	\begin{align}\label{eq: aASC1}
		C_{s, \infty} ^{\rm TS} = T_s \log_2 \left( a_s \rho_b \right) +T_s \sigma_s - \frac{T_s}{\ln 2} \sum_{m = 1}^{M_s} \frac{\xi_m  {\bar F} _{\gamma _{E_s}} \left( \xi_m/\beta_s  \right) /(1+\xi_m )}{(M_s + 1)^2 [L_{M_s + 1}(\xi_m)]^2  \exp(-\xi_m)},
	\end{align}
	\begin{align}\label{eq: aASC2}
		C_{w, \infty} ^{\rm TS} = T_w \log_2 \left( 1+\frac{a_w}{a_s} \right) - \frac{T_w}{\ln 2} \sum_{m = 1}^{M_w} \frac{\pi a_w \sqrt{1 - {\varphi_m}^2} }{\left(a_w \varphi_m+ a_s +1 \right) M_w}  {\bar F} _{\gamma _{E_w}} \left(  \frac{a_w(\varphi_m+1)}{2a_s} \right),
	\end{align}
	\begin{align}\label{eq: aASC3}
		C_{s, \infty} ^{\rm ES} = \log_2 \left( a_s \beta_s \rho_b \right) + \sigma_s - \frac{1}{\ln 2} \sum_{m = 1}^{M_s} \frac{\xi_m  {\bar F} _{\gamma _{E_s}} \left( \xi_m/\beta_s  \right) /(1+\xi_m )}{(M_s + 1)^2 [L_{M_s + 1}(\xi_m)]^2  \exp(-\xi_m)},
	\end{align}
	\begin{align}\label{eq: aASC4}
		C_{w, \infty} ^{\rm ES} = \log_2 \left(1+ \frac{a_w}{a_s} \right) - \frac{1}{\ln 2} \sum_{m = 1}^{M_w} \frac{\pi a_w \sqrt{1 - {\varphi_m}^2} }{\left(a_w \varphi_m+ a_s +1 \right) M_w}  {\bar F} _{\gamma _{E_w}} \left(  \frac{a_w(\varphi_m+1)}{2a_s\beta_w} \right).
	\end{align}
	\hrulefill \vspace*{0pt}
\end{figure*}

\begin{proposition}\label{prop: asymrate TS}
For the TS protocol, the asymptotic ASC in the high-SNR regime can be expressed as \eqref{eq: aASC1} and \eqref{eq: aASC2}, where $\sigma_s = \mathbb{E} [\log_2(H_s)]$.
\end{proposition}
\begin{IEEEproof}
When $\rho_b \to \infty$, \eqref{def: ASC} can be simplified as $	C_{\varepsilon} ^{\rm XS} = \mathbb{E} \left[  C_{U _\varepsilon} ^{\rm XS} \right] - \mathbb{E} \left[C_{E _\varepsilon} ^{\rm XS}   \right] = C_{\varepsilon,max} ^{\rm XS} - \frac{T_\varepsilon}{\ln2} \int_0^{\infty} \frac{{\bar F} _{\gamma _{E_\varepsilon}}(x)  }{1+x} dx$. The term $C_{s,max} ^{\rm TS}$ and $C_{w,max} ^{\rm TS}$ can be expressed as $C_{s,max} ^{\rm TS} \approx \mathbb{E} \left[ T_s \log_2 ( a_s\rho_b H_s ) \right] = T_s \log_2 ( a_s\rho_b) + T_s \sigma_s$ and $C_{w,max} ^{\rm TS} \approx T_w \log_2 \left( 1+ \frac{a_w}{a_s} \right)$, respectively. Then the proposition is proved.
\end{IEEEproof}

\begin{proposition}
For the ES protocol, the asymptotic ASC in the high-SNR regime can be expressed as \eqref{eq: aASC3} and \eqref{eq: aASC4}.
\end{proposition}
\begin{IEEEproof}
The proof is similar to {\bf Proposition \ref{prop: asymrate TS}}.
\end{IEEEproof}

\begin{remark}\label{remark: slope}
In the considered networks, the secrecy slopes of the TS protocol are ${\cal S}_s ^{\rm TS}= T_s$ and ${\cal S}_w ^{\rm TS}= 0$ for the strong LU and the weak LU, respectively. For the ES protocol, the secrecy slopes are ${\cal S}_s ^{\rm ES}= 1$ and ${\cal S}_w ^{\rm ES}= 0$. Therefore, the ES protocol achieves a higher ASC than the TS protocol at a high SNR.
\end{remark}

For clarity, we summarize all results of the secrecy diversity and the secrecy slope for two STAR-RIS protocols in Table \ref{table: sumasmptotic}, where SU represents the strong LU and WU is the weak LU.

\begin{figure*}[t!] 
	\centering
	\subfigure[]{\includegraphics[width=3in]{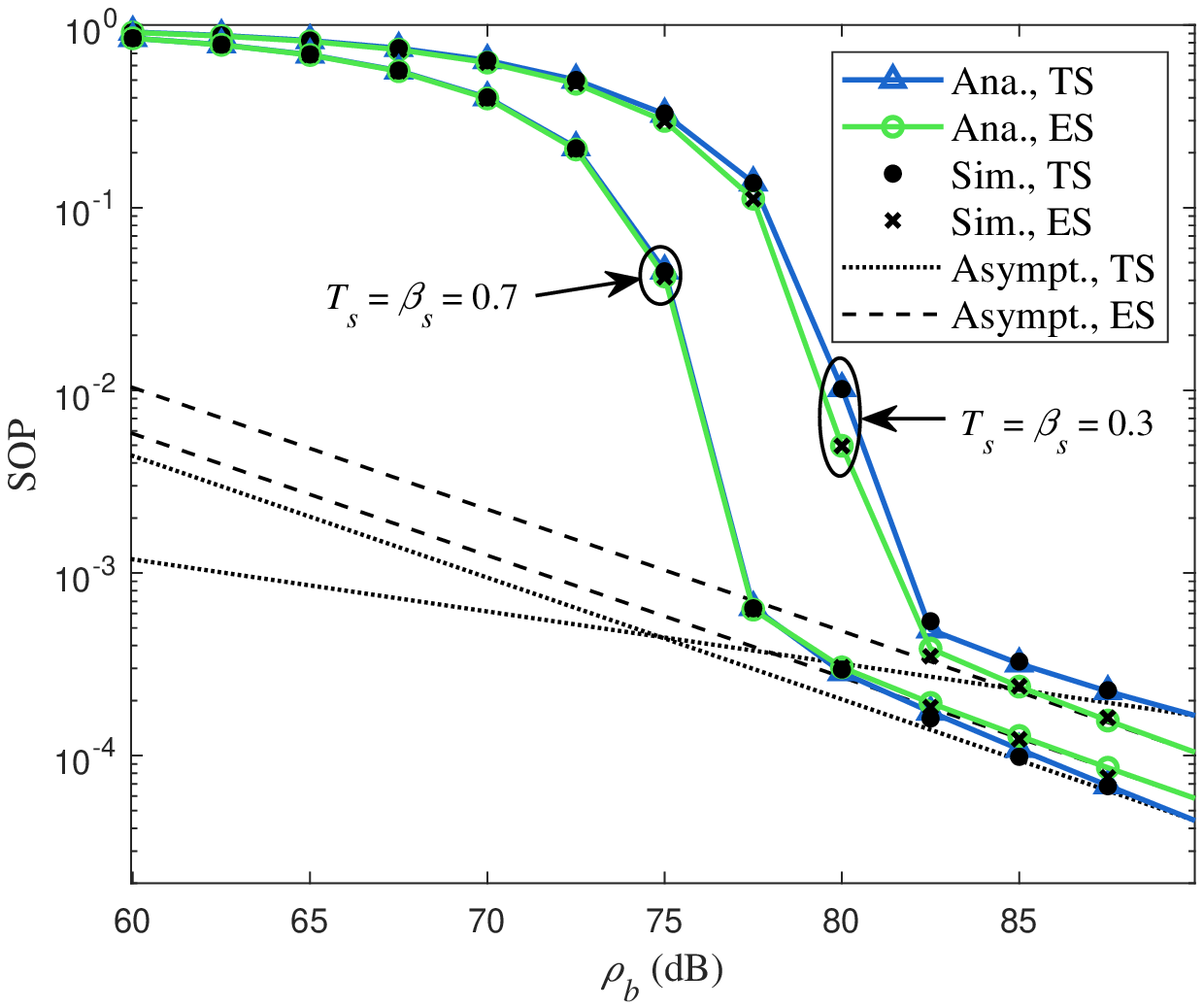}
		\label{1: fig_a}}
	\hfil
	\subfigure[]{\includegraphics[width=3in]{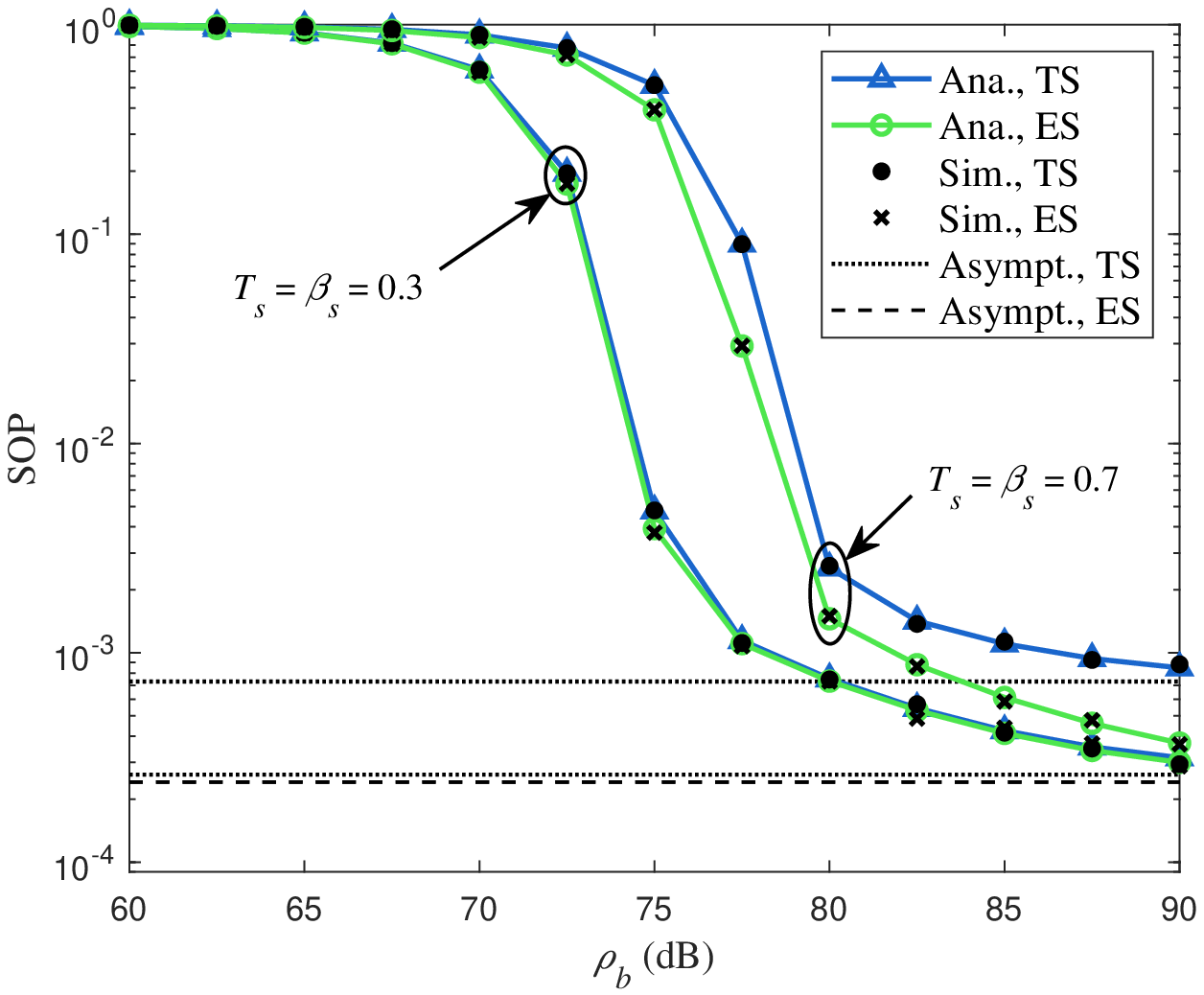}
		\label{1: fig_b}}
	\caption{Validation of the analytical SOP expressions: (a) strong LU; (b) weak LU.
	}
	\label{fig: valid}
	\vspace{-0.3 cm}
\end{figure*}

\begin{figure}[t!]
	\centering
	\includegraphics[width= 3 in]{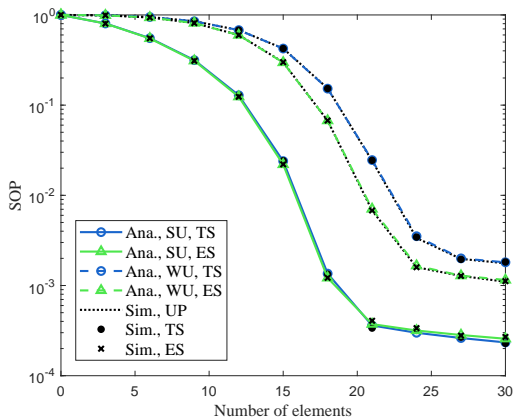}
	\caption{SOP of two NOMA LUs versus the number of STAR-RIS elements with $T_s = \beta_s = 0.7$, where ``SU'' represents strong LU, ``WU'' stands for weak LU, and ``UP'' is the performance for the NOMA LU pair.}\label{fig: numele}
	\vspace{-0.3 cm}
\end{figure}


\vspace{-0.3 cm}
\section{Numerical Results}
In this section, we present the numerical results to demonstrate the performance of STAR-RIS-NOMA. Our theoretical results are validated and then some interesting insights are provided. We mainly focus on the TS and the ES protocols of the STAR-RIS. Unless otherwise stated, the simulation parameters are defined as follows. Consider line-of-sight (LoS) transmissions, the small-scale fading channel is modeled as the cascaded Rician channel \cite{RISPLS1}, and hence we set $\kappa_1 = \kappa_2 = 3$ and $\mu_1 = \mu_2 = 1$. The density of Eves is $\lambda_e = 10^{-4}$ m$^{-2}$. The path loss exponent is $\alpha =3$. The outage threshold for both LUs is $R_s = R_w = 0.1$ bit per channel use. The number of elements on the STAR-RIS is $N = 25$. The radius of the disc area is $R_U = 50$ m. The transmit SNR $\rho_b = 80$ dB and $\rho_e = 50$ dB. The power allocation coefficients for the NOMA LUs are $a_s = 0.3$ and $a_w=0.7$. The complexity-accuracy trade-off parameter $M_u = 30$. The numerical results are verified via Monte Carlo simulations by averaging the obtained performance.

\begin{figure}[t!]
	\centering
	\includegraphics[width= 3 in]{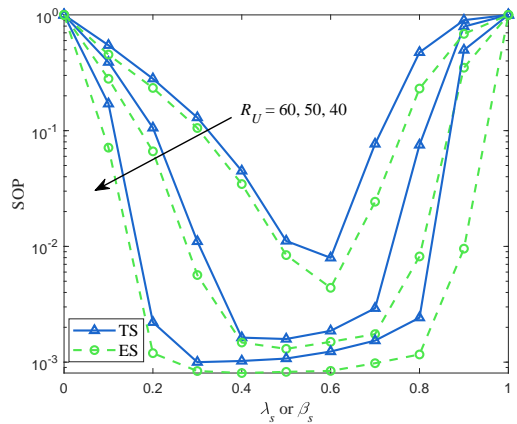}
	\caption{SOP of the NOMA LU pair versus the STAR-RIS mode operation parameter of the strong LU.}\label{fig: parsop}
	\vspace{-0.3 cm}
\end{figure}

\begin{figure*}[t!] 
	\centering
	\subfigure[]{\includegraphics[width=3in]{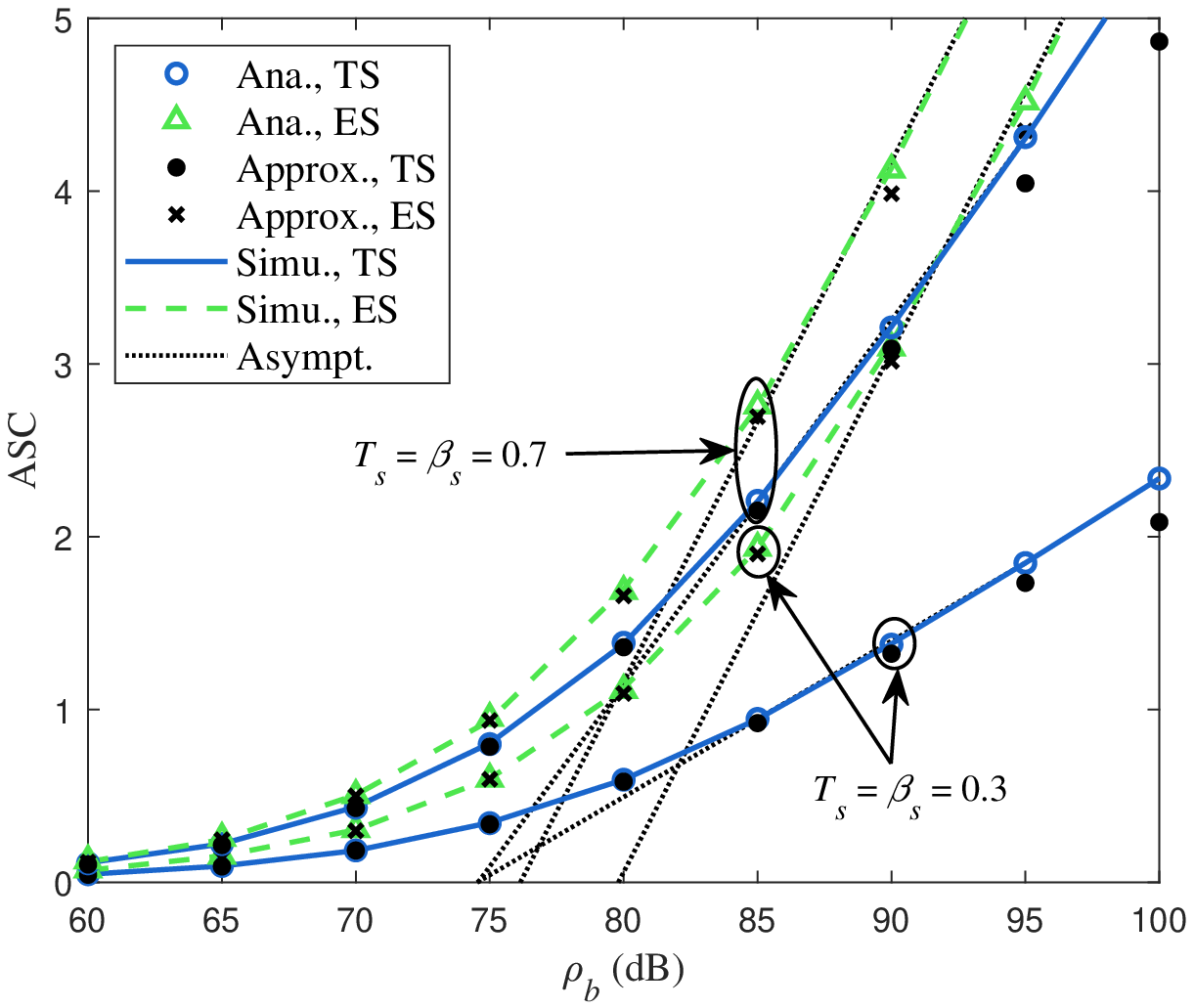}
		\label{3: fig_a}}
	\hfil
	\subfigure[]{\includegraphics[width=3in]{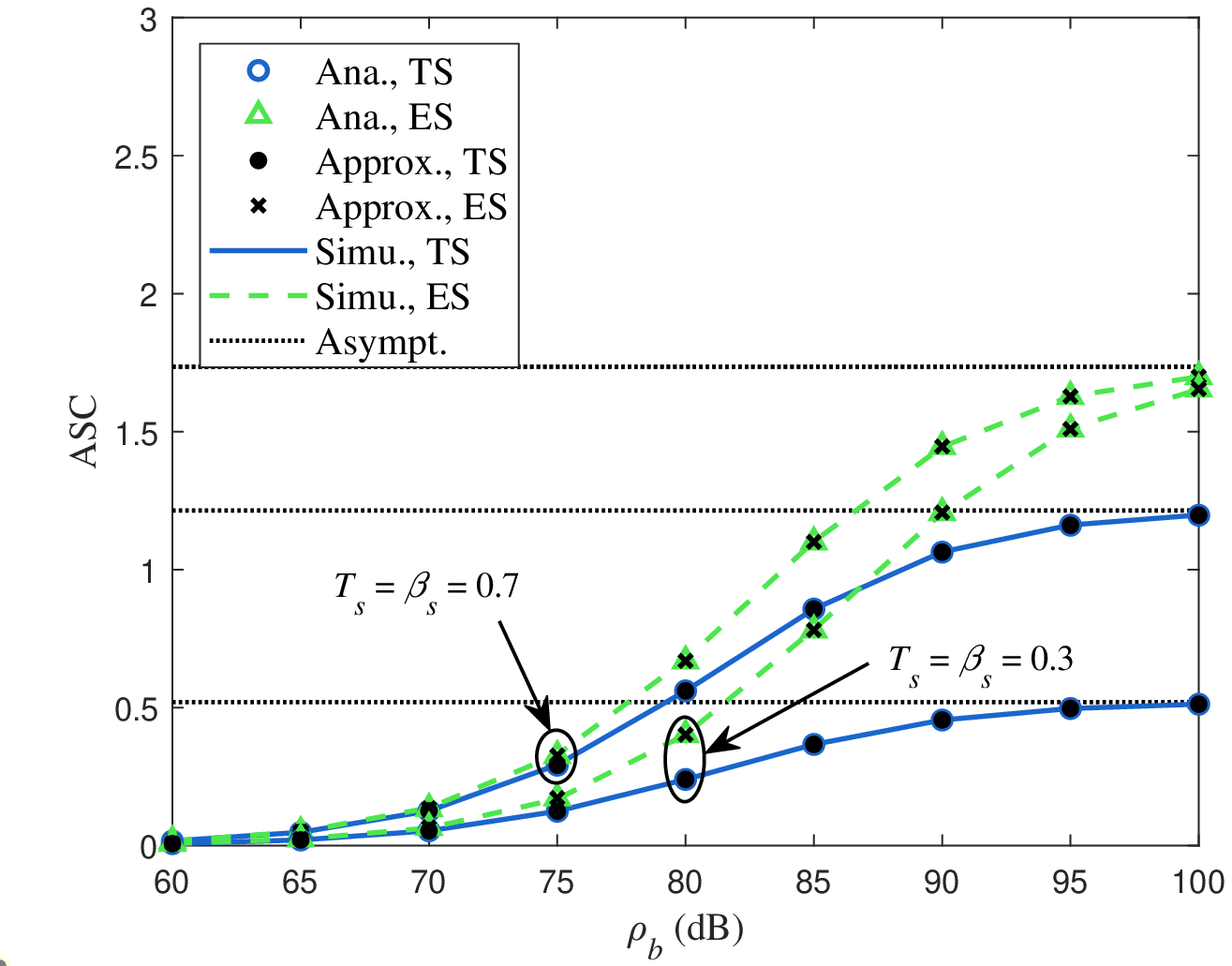}
		\label{3: fig_b}}
	\caption{Validation of the theoretical ASC: (a) strong LU; (b) weak LU.
	}
	\label{fig: validc}
	\vspace{-0.3 cm}
\end{figure*}

\begin{figure}[t!]
	\centering
	\includegraphics[width= 3.1 in]{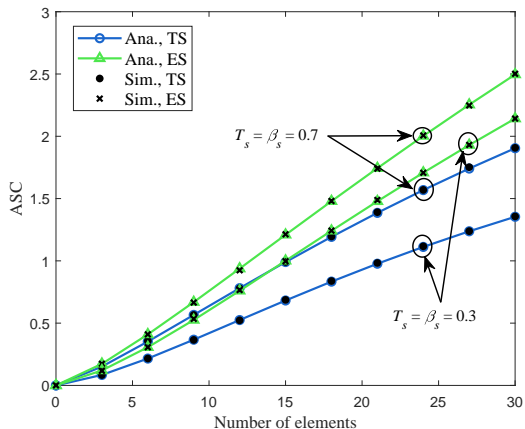}
	\caption{ASC versus the number of STAR-RIS elements.}\label{fig: numeler}
	\vspace{-0.3 cm}
\end{figure}

To illustrate the derived channel statistics in {\bf Lemma \ref{lemma: odered cp CDF}}, Fig. \ref{fig: channelstatistic} plots the CDF of the maximum received SNR of the paired LUs, where the maximum received SNR is the product of transmit SNR $\rho_b$ and the channel power of LU $H_\varepsilon$ for $\varepsilon \in \{s,w\}$. Here we set $\rho_b = 50$ dB for ease of observing different curves. Since {\bf Lemma \ref{lemma: odered cp CDF}} is based on {\bf Lemma \ref{lemma: statistic1}}, the accuracy of the approximation in {\bf Lemma \ref{lemma: statistic1}} is validated. In addition, we observe that the STAR-RIS-aided channel model with a large number of elements has a higher channel power than the model with a few elements. Therefore, the enhanced received SNR at LUs can be obtained by deploying large-scale STAR-RISs.

Fig. \ref{fig: valid} plots the SOP performance of the paired NOMA LUs versus the transmit SNR $\rho_b$. The theoretical curves fit the simulation results quite well and hence {\bf Theorem \ref{theo: NOMA TS}} and {\bf Theorem \ref{theo: NOMA ES}} are validated. Moreover, the asymptotic SOP is presented. As we have discussed in {\bf Corollary \ref{collo: aSOP su}} and {\bf Corollary \ref{collo: aSOP wu}}, the secrecy diversity order of the strong LU is a positive constant related to the path loss exponent while the SOP of the weak LU has an error floor at the high SNR. For the TS protocol, when the time allocation coefficient $T_s < 0.5$, the secrecy diversity order of the strong LU is linear with $T_s$. In this case, the ES protocol has a larger secrecy diversity order than the TS protocol for the strong LU.

Fig. \ref{fig: numele} illustrates the impact of the number of STAR-RIS elements $N$ on the SOP performance. Although the Gamma approximation is adopted to characterize the overall small-scale fading power of STAR-RIS-aided links, it can be observed that the analytical results match the simulation marks even when $N$ is small. Moreover, the SOP of the NOMA LU pair is highly dependent on the worst performance in the paired LUs at a high SNR. With the increase of $N$, the SOP of the NOMA LU pair decreases first but finally keeps at a constant value. This is due to the error floor of the weak LU. Therefore, the increase in the number of elements is able to improve the SOP performance within a certain range, but a large number of elements cannot reduce the error floor of the secure STAR-RIS-NOMA transmission.

Fig. \ref{fig: parsop} plots the SOP of the NOMA LU pair versus the STAR-RIS operation coefficients $T_s$ and $\beta_s$ in the TS protocol and ES protocol, respectively. One can observe that the ES protocol outperforms the TS protocol when $T_s = \beta_s$. Another observation is that there exists an optimal $T_s$ or $\beta_s$ between 0 and 1 to realize the lowest SOP. When $T_s= \beta_s = \{0,1\}$, the secrecy rate of one of the NOMA LUs is zero, and the SOP for the LU pair is one in this case. When $T_s= \beta_s \in (0,1)$, the SOP is smaller than 1 and hence a minimum value exists. Thus a design guideline is provided that the SOP performance can be improved by adjusting the STAR-RIS operation coefficients. Furthermore, the optimal $T_s$ or $\beta_s$ is reduced as $R_U$ increases. Since the SOP of the LU pair mainly depends on the worst performance in two LUs as shown in Fig. \ref{fig: numele}, the optimal $T_s$ or $\beta_s$ is approximated as the crosspoint of the SOP curves of two LUs. With the increase of $R_U$, the gap of best SOP performance of two LUs is bridged and the crosspoint moves to the larger $T_s$ or $\beta_s$.

Fig. \ref{fig: validc} shows the ASC of two NOMA LUs versus the transmit SNR $\rho_b$. The analytical results are from {\bf Theorem \ref{theo: cap TS}} and {\bf Theorem \ref{theo: cap ES}}. The approximation results are obtained in {\bf Corollary \ref{coro: cap TS}} and {\bf Corollary \ref{coro: cap ES}}. We can observe that the approximation results of the weak LU fit the simulation curves well while there is a small performance gap for the strong LU especially at a high SNR. This is because the Chebyshev-Gauss quadrature is accurate even with a small $M_u$. However, the parameter $M_u$ is not large enough to ensure the accuracy of the Gauss–Laguerre quadrature. In Fig. \ref{3: fig_a}, the secrecy slope for the strong LU in the TS protocol is smaller than the ES protocol as discussed in {\bf Remark \ref{remark: slope}}. In Fig. \ref{3: fig_b}, with the increase of the transmit SNR $\rho_b$, the ASC of the weak LU achieves an upper bound. In the ES protocol, the capacity upper bounds for two NOMA LUs are the same. Moreover, the ES protocol has a higher ASC performance than the TS protocol since extra flexibility from the space is utilized by the ES protocol.

\begin{figure}[t!]
	\centering
	\includegraphics[width= 3 in]{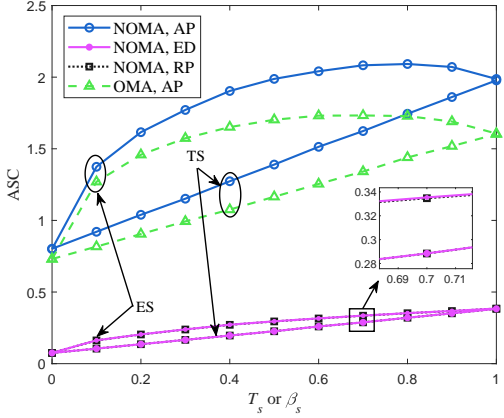}
	\caption{ASC versus the STAR-RIS mode operation parameter of the strong LU, where ``AP'' represents the aligned phase scheme in our analysis, ``ED'' denotes the Eve depression scheme, and ``RP'' is the random phase.}\label{fig: capacompare}
	\vspace{-0.3 cm}
\end{figure}

Fig. \ref{fig: numeler} plots the ASC of the NOMA LU pair versus the number of STAR-RIS elements for two protocols. Different from the observation in the SOP performance, the ASC keeps growing linearly as the number of elements increases. The reason is that the achievable rate of the strong LU can be improved by strengthening the RIS-assisted channel. Therefore, a high ASC can be achieved by employing a STAR-RIS with a large number of elements.

Fig. \ref{fig: capacompare} compares the ASC performance among different scenarios under different STAR-RIS mode operation parameter values. We observe that the ASC of NOMA obtains a significant improvement over OMA. This illustrates the efficiency of adopting the NOMA scheme in STAR-RIS-aided systems. To show the efficiency of the phase aligned scheme at the paired NOMA LUs, we plot curves of the Eve depression scheme and the random phase setup for comparison. In the Eve depression scheme, the STAR-RIS suppresses the eavesdropping of the most detrimental Eve and reduces its capacity to zero as in \cite{RISPLS0}. We can observe that the phase aligned scheme is able to improve the secrecy rate performance remarkably, while the enhancement of the Eve depression scheme is negligible in the considered multi-Eve system. Moreover, when we set $T_s = \beta_s$, the rate performance of the ES protocol always outperforms the TS protocol due to the extra degree of freedom in the space domain. By adjusting $T_s$ or $\beta_s$, the highest ASC can be achieved. One should be noted that the ASC of the TS protocol is linear with $T_s$ ({\bf Remark \ref{remark: Linearrate TS}}), and hence $T_s = 1$ is optimal. However, the optimal $\beta_s$ depends on the system settings.

\section{Conclusion}
In this paper, the PLS of the STAR-RIS-NOMA has been investigated. The stochastic geometry based tool has been utilized to model the random locations of NOMA LUs and the Eves. Considering the TS protocol and the ES protocol, we have derived the analytical expressions of the SOP and the ASC when the SIC order of the NOMA LUs is based on the channel gains. In the high SNR regime, the asymptotic secrecy performance has been obtained. The analytical results have revealed that the error floor exists for the SOP in the secure STAR-RIS-NOMA transmission. The numerical results have provided design guidelines for the considered system: 1) the optimal secrecy performance can be achieved by adjusting the mode operation parameters of the STAR-RIS; 2) the ES protocol has a better secrecy performance than the TS protocol; 3) the STAR-RIS with a large number of elements can be employed for the high ASC.

\section*{Appendix~A: Proof of Lemma~\ref{lemma: odered cp CDF}}
\label{Appendix:A}
\renewcommand{\theequation}{A.\arabic{equation}}
\setcounter{equation}{0}

In this work, the overall channel power consists of path loss and small-scale fading. We denote $H_{u} = X Y$, where $X = |h_\varepsilon|^2$ and $Y = A_Ld^{-\alpha}$ represent the power of small-scale fading and path loss at the LU, respectively. According to \eqref{eq: sfading user}, the CDF of the small-scale fading $X$ is
\begin{align}
	F_X(x) = \frac{\gamma\left(k_r, x/\theta_r \right)}{\Gamma(k_r)}.
\end{align}

Noticed that the locations of LUs obey a HPPP in the disc area, the PDF of the path loss $Y$ is given by
\begin{align}
	f_Y(x) = 
	\left\{ 
	\begin{array}{l}
		\frac{2{A_L}^{2/\alpha}}{\alpha{R_U}^2}x^{-2/\alpha-1} ,x > {A_L} {R_U}^{-\alpha} \\
		0,x \le {A_L}{R_U}^{-\alpha}. \\
	\end{array} \right.
\end{align}

For an arbitrary LU in ${\bf \Phi}_u$, we can formulate the CDF of the channel power $H_{u}$ as follows 
\begin{align}\label{eq: A.3}
	\hat{F}_{H_{u}}(x) &= \int_{0}^\infty F_X(\frac{x}{y}) f_Y(y)dy \nonumber \\
	& \overset{(a)}{=}  \frac{2}{{R_U}^2}  \int_0^{R_U} \frac{\gamma\left(k_r, \frac{x r^{\alpha}}{ A_L \theta_r} \right)}{\Gamma(k_r)} r dr,
\end{align}
where $(a)$ is from the change of variable $r = (y/A_L)^{-1/\alpha}$. By employing the meijer G-function of lower incomplete Gamma function, we rewrite $\hat{F}_{H_{u}}(x)$ as
\begin{align}\label{eq: A.4}
	\hat{F}_{H_{u}}(x) & = \frac{2}{{R_U}^2 \Gamma(k_r)} \int_0^{R_U} r G_{1,1}^{1,2} \left( \frac{xr^\alpha}{A_L \theta_r}  \left| \begin{matrix}
		1 \\ k_r,0
	\end{matrix} \right. \right) dr \nonumber \\
	& \overset{(b)}{=} \frac{\delta}{\Gamma (k_r)} G_{2,3}^{1,2} \left( \frac{{R_U}^\alpha x}{A_L \theta_r}  \left| \begin{matrix}
		1-\delta,1 \\ k_r,0,-\delta
	\end{matrix} \right. \right),
\end{align}
where $(b)$ is obtained by utilizing \cite[eq. (7.811.2)]{Intetable}.


For the LU pair, according to order statistics theory \cite{statistorder}, if total of $K$ LUs have the same statitical channel characteristic, the ordered CDF of the channel power of the $l$th weakest LU is given by
\begin{align}\label{eq: A.5}
	F_l(x) &= \sum_{k = l}^{K} \binom{K}{k} [{\hat F}_{H_{u}}(x)]^{k} [1-{\hat F}_{H_{u}}(x)]^{K-k}.
\end{align}


By substuting \eqref{eq: A.4} into \eqref{eq: A.5}, this lemma is proved.

\renewcommand{\theequation}{B.\arabic{equation}}
\setcounter{equation}{0}

\begin{figure*}[t!]
	\normalsize
	\begin{align}\label{eq: B.1}
		{\cal L}_{f_{\Delta_n}} (\omega) &= \frac{2 \sqrt {\phi_1 \phi_2}}{e^{\mu_1 \kappa_1 + \mu_2 \kappa_2}} \sum_{q=0}^{\infty} \sum_{t=0}^{\infty} \rho_{q,t} \int_0^{\infty} e^{-\omega x} G_{0,2}^{2,0} \left( \phi_1 \phi_2x^2 \left| q+\mu_1-\frac{1}{2}, t+\mu_2-\frac{1}{2} \right. \right) dx \nonumber\\ 
		&= \frac{2 \sqrt {\phi_1 \phi_2}}{\sqrt{\pi} e^{\mu_1 \kappa_1 + \mu_2 \kappa_2}} \sum_{q =0}^{\infty} \sum_{t=0}^{\infty} \frac{\rho_{q,t}}{\omega} \underbrace{G_{2,2}^{2,2} \left( \frac{4\phi_1 \phi_2}{\omega^2}  \left| \begin{matrix}
				0,\frac{1}{2} \\ q+\mu_1-\frac{1}{2}, t+\mu_2-\frac{1}{2}
			\end{matrix} \right. \right)} _{J_{q,t}} .
	\end{align}
	\hrulefill \vspace*{0pt}
\end{figure*}

\section*{Appendix~B: Proof of Lemma~\ref{lemma: exact order}}
\label{Appendix:B}
To obtain the accurate asymptotic performance, we consider the accurate expression rather than the Gamma approximation for the small-scale fading power $|h_u|^2$. Note that all channels of the STAR-RIS are independent, we employ the convolution method to derive the distribution of the overall small-scale fading power $|h_u|^2$. Based on \eqref{eq: doublefadn pdf}, we first calculate the Laplace transform of $f_{\Delta_n}(x)$ as in \eqref{eq: B.1}.
We denote $J_{q,t} = G_{2,2}^{2,2} \left( \frac{4\phi_1 \phi_2}{\omega^2}  \left| \begin{matrix}
		{\bf a} \\ {\bf b}
\end{matrix} \right. \right)$. By utilizing the relationship between the meijer G-function and the generalized hypergeometric function, $J_{q,t}$ is rewritten as
\begin{align}
	J_{q,t} = \sum_{m=1}^2 \left( \frac{4\phi_1 \phi_2}{\omega^2} \right)^{b_m} \underbrace{{}_2F_1 \left( \frac{4\phi_1 \phi_2}{\omega^2}  \left| \begin{matrix}
			1+b_m-{\bf a} \\ 1+b_m-{\bf b}
		\end{matrix} \right. \right)}_{H_1} \nonumber \\
	\times \prod_{i =1}^{2} \Gamma(b_i - b_m) \prod_{i =1}^{2} \Gamma(1+b_m - a_i) .
\end{align}
	
We observe that \eqref{eq: B.1} is so complicated that it is difficult to obtain the tractable expression of the inverse Laplace transform for $\prod_{n=1}^N {\cal L}_{f_{\Delta_n}} (\omega)$. Therefore, we consider the case $\omega \to \infty$ to calculate the PDF of the overall channel gain near 0. When $\omega \to \infty$, $H_1 \to 1$ holds. We only keep the dominant item in \eqref{eq: B.1}, i.e., the item with $r=t=0$, and we have
\begin{align}
	{\cal L}_{f_{\Delta_n}}^{\infty} (\omega) = A_u \omega^{-2 \hat{\mu}},
\end{align}
where $\hat{\mu} = \min \{\mu_1, \mu_2\}$ and $A_u$ is a constant unrelated to $\omega$. Since all channels of the STAR-RIS are i.i.d., the Laplace transform of the PDF for the overall small-scale fading gain $|h_u|$ is
\begin{align}
	{\cal L}_{f_{|h_u|}}^{\infty} (\omega) = {A_u}^N \omega^{-2 \hat{\mu}N}.
\end{align}
	
We are able to obtain the PDF of $|h_u|$ by conducting the inverse Laplace transform of ${\cal L}_{f_{|h_u|}}^{\infty} (x)$, which is given by
\begin{align}
	f_{|h_n|}^{0^+} (x) = \frac{{A_u}^N}{\Gamma(2 \hat{\mu} N)} x^{2 \hat{\mu}N -1}.
\end{align}
	
Then we obtain the CDF of $|h_u|$ as follows
\begin{align}
	F_{|h_u|}^{0^+} (x) = \frac{{A_u}^N}{\Gamma(2 \hat{\mu} N + 1)} x^{2 \hat{\mu}N}.
\end{align}
	
Afterwards, the CDF of the overall small-scale fading power $|h_u|^2$ can be easily calculated, which is given by
\begin{align}
	F_{|h_{u}|^2}^{0^+}(x) = \frac{{A_u}^N}{\Gamma(2 \hat{\mu} N + 1)} x^{\hat{\mu}N} ,
\end{align}
where $\hat{\mu} = \min \{\mu_1, \mu_2\}$ and  $A_u = \frac{ K_u \rho_{0,0} \left(\phi_1 \phi_2\right)^{\hat{\mu}} \Gamma(|\mu_1-\mu_2|) \Gamma(\frac{1}{2}+\hat{\mu}) \Gamma(\hat{\mu}) }{\sqrt{\pi}e^{\mu_1 \kappa_1 + \mu_2 \kappa_2}}$. $K_u = 2$ when $\mu_1 = \mu_2 $; otherwise, $K_u = 1$. Based on \eqref{eq: A.3}, we are able to obtain the CDF of the asymptotic unordered channel $H_u$
\begin{align}
	\hat{F}_{H_{u}}^{0^+}(x) = \frac{2}{{R_U}^2} \int_0^{R_U}\frac{{A_u}^N (x/A_L)^{\hat{\mu}N}}{\Gamma(2 \hat{\mu} N + 1)} r^{\alpha\hat{\mu}N+1} dr = L_u x^{\hat{\mu}N}.
\end{align}

Then the proof is completed.

\section*{Appendix~C: Proof of Theorem~\ref{theo: cap TS}}
\label{Appendix:C}
\renewcommand{\theequation}{C.\arabic{equation}}
\setcounter{equation}{0}

Based on the definition in \eqref{def: ASC}, the ASC for the strong LU is expressed as
\begin{align}\label{eq: C.1}
	C_{s} ^{\rm TS} &= \int_0 ^\infty \int_0 ^x T_s\log_2 \frac{(1+x)}{(1+y)} f_{\gamma_{U_s} ^{\rm TS}}(x) f_{\gamma_{E_s} ^{\rm TS}}(y) dy dx \nonumber \\
	& = \int_0 ^\infty \int_0 ^\infty T_s \log_2(1+x) f_{\gamma_{U_s} ^{\rm TS}}(x) f_{\gamma_{E_s} ^{\rm TS}}(y) dy dx \nonumber \\ 
	& - \int_0 ^\infty  T_s\log_2(1+x)f_{\gamma_{U_s} ^{\rm TS}}(x) \bar{F}_{\gamma_{E_s} ^{\rm TS}}(x) dx \nonumber \\
	& - \int_0 ^\infty  T_s\log_2(1+x)  \bar{F} _{\gamma_{U_s} ^{\rm TS}}(x) f_{\gamma_{E_s} ^{\rm TS}}(x) dx \nonumber \\
	& = \underbrace{\frac{T_s}{\ln2} \int_0^{\infty} \frac{{\bar F} _{\gamma _{U_s}}(x)  }{1+x} dx}_{C_{s,\max} ^{\rm TS}} - \underbrace{\frac{T_s}{\ln2} \int_0^{\infty} \frac{{\bar F} _{\gamma _{U_s}}(x) {\bar F} _{\gamma _{E_s}}(x)  }{1+x} dx}_{C_{s,{\rm loss}} ^{\rm TS}},
\end{align}
where $C_{s,\max} ^{\rm TS}$ is the ASC without eavesdropping, $C_{s,{\rm loss}} ^{\rm TS}$ is the ASC loss due to the most detrimental Eve. 

For the weak LU, the ASC is zero when $a_w - \gamma_{U_s} ^{\rm TS} a_s \le 0$. Thus the ASC is given by
\begin{align}\label{eq: C.2}
	C_{w} ^{\rm TS} = &\underbrace{\frac{T_w}{\ln2} \int_0^{\frac{a_w}{a_s}} \frac{{\bar F} _{\gamma _{U_w}}(x)  }{1+x} dx}_{C_{w,\max} ^{\rm TS}} \nonumber \\
	&- \underbrace{\frac{T_w}{\ln2} \int_0^{\frac{a_w}{a_s}} \frac{{\bar F} _{\gamma _{U_w}}(x) {\bar F} _{\gamma _{E_w}}(x)  }{1+x} dx}_{C_{w,{\rm loss}} ^{\rm TS}}.
\end{align}

Then the proof is completed.

\bibliographystyle{IEEEtran}
\bibliography{reference}

\end{document}